\documentclass[journal]{IEEEtran}
\usepackage{amsmath,amsfonts}
\usepackage{algorithmic}
\usepackage{algorithm}
\usepackage{array}
\usepackage[caption=false,font=normalsize,labelfont=sf,textfont=sf]{subfig}
\usepackage{textcomp}
\usepackage{stfloats}
\usepackage{url}
\usepackage{verbatim}
\usepackage{graphicx}
\usepackage{cite}
\usepackage{hyperref}
\usepackage{booktabs}
\usepackage{multirow}
\usepackage{enumitem}

\usepackage{color}

\begin{document}

\title{ASiM: Modeling and Analyzing Inference Accuracy of SRAM-Based Analog CiM Circuits}

\author{Wenlun Zhang~\IEEEmembership{Graduate Student Member,~IEEE,}, Shimpei Ando,\\Yung-Chin Chen, and Kentaro Yoshioka~\IEEEmembership{Member,~IEEE,}
        % <-this % stops a space
\thanks{The authors are with the Department of Electronics and Electrical Engineering, Keio University, Kanagawa 223-8522, Japan (e-mail: wenlun\_zhang@keio.jp). This research was supported in part by the JST CREST JPMJCR21D2, JSPS Kakenhi 23H00467, Futaba Foundation, Asahi Glass Foundation, Telecommunications Advancement Foundation, and the KLL Ph.D. Program Research Grant. We also thank all anonymous reviewers and area chairs for their constructive feedback.
}% <-this % stops a space
}

\markboth{arXiv Preprint, August~2025}%
{Shell \MakeLowercase{\textit{et al.}}: A Sample Article Using IEEEtran.cls for IEEE Journals}


\maketitle

\begin{abstract}
SRAM-based Analog Compute-in-Memory (ACiM) demonstrates promising energy efficiency for Deep Neural Network (DNN) processing. Nevertheless, efforts to optimize efficiency frequently compromise accuracy, and this trade-off remains insufficiently studied due to the difficulty of performing full-system validation. Specifically, existing simulation tools rarely target SRAM-based ACiM and exhibit inconsistent accuracy predictions, highlighting the need for a standardized, SRAM CiM circuit-aware evaluation methodology. This paper presents ASiM, a simulation framework for evaluating inference accuracy in SRAM-based ACiM systems. ASiM captures critical effects in SRAM based analog compute in memory systems, such as ADC quantization, bit parallel encoding, and analog noise, which must be modeled with high fidelity due to their distinct behavior in charge domain architectures compared to other memory technologies. ASiM supports a wide range of modern DNN workloads, including CNN and Transformer-based models such as ViT, and scales to large-scale tasks like ImageNet classification. Our results indicate that bit-parallel encoding can improve energy efficiency with only modest accuracy degradation; however, even 1 LSB of analog noise can significantly impair inference performance, particularly in complex tasks such as ImageNet. To address this, we explore hybrid analog-digital execution and majority voting schemes, both of which enhance robustness without negating energy savings. ASiM bridges the gap between hardware design and inference performance, offering actionable insights for energy-efficient, high-accuracy ACiM deployment. \textbf{Code is available at} \href{https://github.com/Keio-CSG/ASiM}{\textit{https://github.com/Keio-CSG/ASiM}}
\end{abstract}


\section{Introduction}
\label{Chapt_Introduction}

\IEEEPARstart{D}{eep} learning has brought revolutionary changes to a wide range of fields, not only in computer vision and natural language processing but also complex tasks spanning different modalities, delivering state-of-the-art performance. Recent advances have shown a scaling law in deep learning models, where performance consistently improves as the size of the model increases~\cite{LLM_Scaling}. However, executing these large-scale models in traditional von Neumann architectures leads to substantial energy consumption and latency~\cite{Energy_Problem}, which poses critical challenges, particularly for edge devices with limited resources. Compute-in-Memory (CiM) has emerged as a novel paradigm to overcome these limitations by performing computations directly within the memory arrays ~\cite{IMC_Advances_Prospects}. By reducing data movement between logic and memory, CiM architectures can achieve significant improvements in both energy efficiency and throughput.

Among CiM architectures, Analog Compute-in-Memory (ACiM) offers high computational density and energy efficiency by leveraging the analog properties of memory cells to perform Multiply-and-ACcumulate (MAC) operations directly in the analog domain~\cite{Analog_or_Digital}. SRAM-based ACiM, in particular, has attracted attention due to its compatibility with standard CMOS processes and reliable charge-domain operation~\cite{64-Tile}. In recent years, state-of-the-art ACiM designs have tended to compress input activations and perform MAC operations in a bit-parallel fashion while lowering ADC bit precision to achieve improved energy efficiency~\cite{MACC-SRAM,In-SRAM-ADC_ACIM,PIMCA,One-Shot_AD_MAC,PICO-RAM}. These techniques, while beneficial in some respects, exacerbate quantization errors and make it difficult to preserve inference accuracy, highlighting the importance of accurate and thorough evaluation.

Despite of successive improvement on energy efficiency of ACiM designs, discussion regarding inference accuracy remains limited. This gap of understanding inference accuracy is particularly concerning by the challenges of validating ACiM circuits, because the massive scale of modern DNNs makes full hardware validation increasingly difficult. A more practical approach is partial validation, where researchers measure comprehensive characteristics of ACiM circuits and integrate them into software simulations to evaluate the end-to-end inference accuracy~\cite{Programmable_Accelerator_ISSCC, C3SRAM,MACC-SRAM,In-SRAM-ADC_ACIM}. Although this approach theoretically yields simulation results comparable to full hardware measurements, modeling ACiM circuits presents numerous latent challenges that can lead to misleading accuracy assessments. 

We examined several existing simulation frameworks aimed at assessing inference accuracy and observed significant variation in their predictions. These discrepancies suggest that current tools may struggle to provide consistent guidance for design validation or silicon correlation. A key contributing factor is the lack of standardized, open-source simulators tailored specifically to SRAM-based ACiM. The inconsistent predictions produced by different tools highlight a key research question: \textbf{What modeling components are essential for accurate inference evaluation in SRAM-based ACiM systems, and how can we ensure reliable, consistent predictions across diverse design scenarios?} Addressing this gap would support the development of standardized evaluation practices and facilitate the broader adoption of ACiM architectures.

In this paper, we introduce ASiM, an accurate simulation framework that bridges low-level circuit behavior and high-level inference performance in SRAM-based ACiM systems. ASiM models essential circuit effects such as ADC quantization, bit-parallel encoding, and analog noise, enabling accurate evaluation of how hardware design choices impact DNN inference accuracy. 
Designed for practical use, ASiM integrates seamlessly with PyTorch as a plug-and-play tool, providing both modeling fidelity and ease of deployment. It supports SRAM ACiM specific architectural innovations, including bit-parallel execution, and is compatible with modern DNNs such as Transformers. Using ASiM, we conduct a detailed analysis of how various circuit-level factors influence inference accuracy. The main \textbf{findings} are as follows:

\begin{itemize}[leftmargin=*]
    \item \textbf{Analog noise within ACiM circuits severely impact inference accuracy:} Our analysis reveals that even minor ADC readout errors of 1 LSB can propagate through the network and cause significant accuracy degradation.
    \item \textbf{Complex tasks demand stricter hardware design constraints:} Inference on tasks like ImageNet is substantially more sensitive to circuit-level nonidealities than simpler benchmarks such as CIFAR-10. Maintaining accuracy on such tasks requires tighter design margins or the adoption of more robust approaches like Hybrid CiM (HCiM).
    \item \textbf{Bit-parallel ACiM improves energy efficiency but increases sensitivity to noise:} While this architecture tolerates moderate quantization noise and achieves substantial energy savings, it is more susceptible to analog noise, necessitating stricter design standards to ensure inference reliability.
\end{itemize}

\noindent This paper makes the following \textbf{contributions}:

\begin{itemize}[leftmargin=*]
    \item We develop \textbf{ASiM}, an open-source simulation framework for systematic inference accuracy evaluation in SRAM-based ACiM systems. ASiM integrates with PyTorch and supports modern DNN architectures, including Transformers.
    \item With ASiM, we analyze the impact of key circuit-level design factors on inference accuracy and uncover how specific noise sources degrade model performance, revealing sensitivity issues often overlooked in prior work.
    \item Building on our findings, we provide practical design guidelines and solutions to improve ACiM inference accuracy, enhancing its practicality for real-world applications.
\end{itemize}

The remainder of this paper is organized as follows. Section~\ref{Chapt_Preliminaries} reviews related works and provides preliminary concepts on ACiM and inference evaluation metrics. Section~\ref{Chapt_Framework} describes the detailed implementation and modeling methodology of the ASiM framework. Section~\ref{Chapt_Analysis} presents a comprehensive analysis of how each ACiM design factor impacts DNN inference accuracy and highlights fundamental limitations. Section~\ref{Chapt_Accurate_Inference} provides simple yet effective solutions to improve ACiM inference accuracy. Section~\ref{Chapt_Discussion} offers a comprehensive discussion of the ASiM framework alongside state-of-the-art ACiM designs. Finally, Section~\ref{Chapt_Conclusion} concludes the paper.

\section{Preliminaries}
\label{Chapt_Preliminaries}

\subsection{Charge Domain ACiM using SRAM}

CiM accelerators break down multi-bit matrix multiplications into bit-wise MAC operations~\cite{Bit-Flexible_CIM}, as described in Eq.~\ref{Eq_MAC}:

\begin{equation}
O = \mathbf{x} \cdot \mathbf{w} = \sum_{p = 0}^{P - 1} \sum_{q = 0}^{Q - 1} (-1)^{k} {2}^{p + q} \sum_{n = 0}^{N - 1} x_{n}[p] w_{n}[q].
\label{Eq_MAC}
\end{equation}

\noindent Here, $\mathbf{x}$ and $\mathbf{w}$ denote the N-dimensional input activation vector and weight vector, respectively, both represented in integer format with bit widths of P and Q. The factor $k \in \{0, 1\}$ is used to manage negative conditions in 2's complement for each bit-wise MAC cycle. In SRAM-based charge domain ACiM, weight bits are stored directly in memory cells to reduce frequent memory accesses, while activations are continuously refreshed and broadcast across the ACiM array. The resulting Dot Products (DPs) are temporarily stored in local capacitors; the accumulated charge is then shared or coupled along a Compute Bit-Line (CBL), and the voltage is finally read out using ADCs. 

\begin{figure}[htbp]
    \centering
    \subfloat[]{%
        \includegraphics[width=0.98\linewidth]{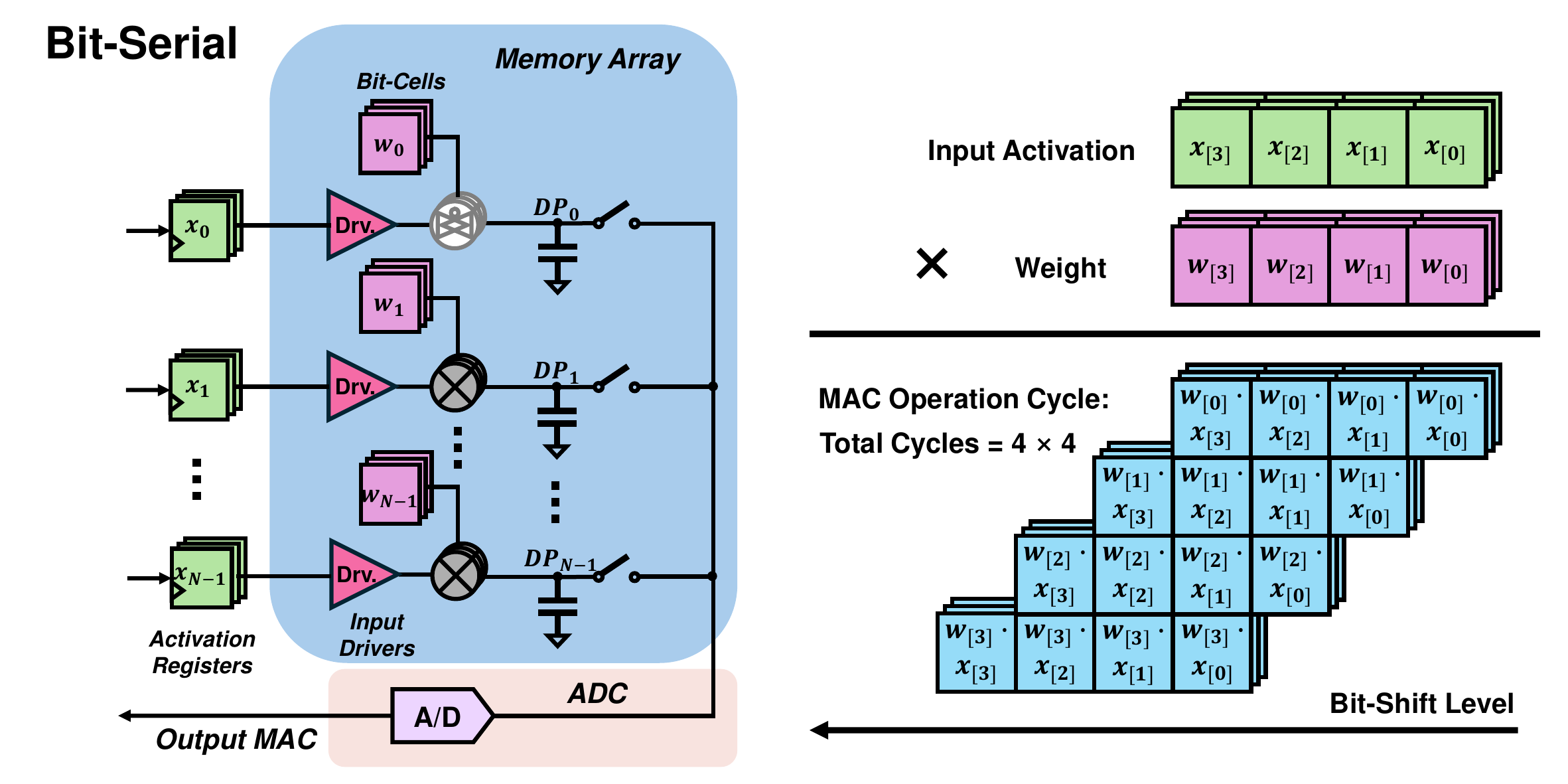}
        \label{Fig_ACiM_a}
    }\\

    \subfloat[]{%
        \includegraphics[width=0.98\linewidth]{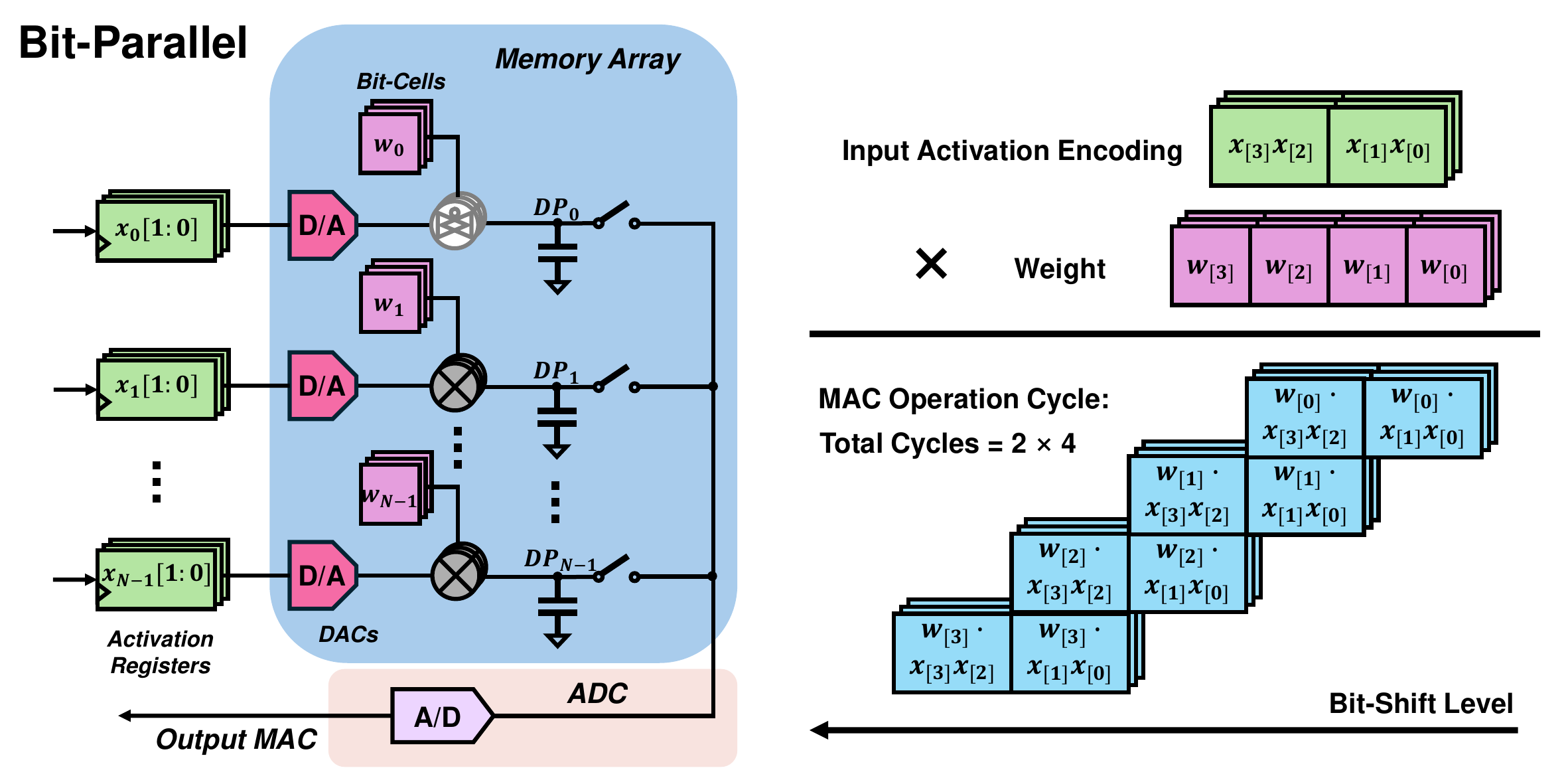}
        \label{Fig_ACiM_b}
    }%
    
    \caption{Fundamental architecture and MAC operation of ACiM. (a) Bit-serial ACiM. (b) Bit-parallel ACiM.}
    \label{Fig_ACiM}
\end{figure}

Early version of bit-serial ACiM operates by broadcasting activations bit by bit, causing local capacitors to undergo either a full charge or discharge during each cycle. This effectively decomposes a matrix multiplication of $P$ and $Q$ bits into $P \times Q$ binary MAC cycles, as shown in Fig.~\ref{Fig_ACiM}(a). 
Progressing from bit-serial ACiM, the bit-parallel ACiM encodes input activations into $y$-bit compression via DACs, which allows multi-step DP voltage levels in the local capacitors and reduces the MAC cycles approximately to $P / y \times Q$~\cite{PICO-RAM}, as illustrated in Fig.~\ref{Fig_ACiM}(b). By minimizing the number of MAC cycles, bit-parallel ACiM significantly enhances energy efficiency. 

However, compressing more activation bits into a single ADC readout can potentially degrade inference accuracy, as it introduces higher quantization noise than the traditional bit-serial scheme. Additionally, recent designs often reduce ADC precision to achieve maximum energy efficiency, further amplifying quantization noise, which may compromise DNN inference. Under these conditions, the intrinsic nonidealities of analog computations in ACiM, such as noise and signal distortion, become increasingly problematic. This creates additional complexities in determining optimal design policies. Despite successive advancements in ACiM architectures, there remains a lack of comprehensive analysis on how these factors affect inference accuracy. To fill this critical gap, we have developed ASiM, a comprehensive simulation framework, to analyze these effects and highlight the bottlenecks in charge domain ACiM, providing insights for future ACiM design improvements.

\subsection{Simulation Frameworks for ACiM}

Full DNN inference validation on ACiM prototypes requires significant resources. As an alternative, various simulation frameworks have been developed to model ACiM performance across memory types. These tools aid both pre-silicon design exploration and post-silicon performance evaluation. Some emphasize circuit-level performance modeling~\cite{NeuroSim,SySCIM,PIMSIM-NN}, while others incorporate analog nonidealities to assess inference accuracy~\cite{aihwkit,MemTorch,eF2lowSim,Fast_Accurate_Sim}. Frameworks such as X-PIM~\cite{X-PIM} aim to address both.

However, most existing frameworks are optimized for non-volatile memories like RRAM and Flash, whose noise characteristics differ from SRAM. In contrast, ASiM adheres to SRAM-specific modeling principles, using binary cycle decomposition and fine-grained modeling of MAC outputs to reflect real fabricated circuit behavior. Implemented within PyTorch, ASiM supports recent ACiM features including bit-parallel operations and Transformer-based models, enabling flexible, accurate performance evaluation across diverse workloads.

\subsection{Evaluation Metrics for Inference Accuracy}

To quantify inference degradation from ACiM execution, metrics such as Signal to Quantization Noise Ratio (SQNR) and Compute Signal to Noise Ratio (CSNR) are commonly used~\cite{Accuracy_Efficiency_Trade_Off}. SQNR captures noise from quantization alone, while CSNR accounts for additional analog effects. The original definitions of SQNR and CSNR are given in~\cite{CSNR_Metrics} as:

\begin{equation}
\text{SQNR} = \frac{{\sigma}^{2}_{y}}{{\sigma}^{2}_{y_i}} + \frac{{\sigma}^{2}_{y}}{{\sigma}^{2}_{y_o}},
\label{Eq_SQNR_Def}
\end{equation}

\begin{equation}
\text{CSNR} = \frac{{\sigma}^{2}_{y}}{{\sigma}^{2}_{y_i}} + \frac{{\sigma}^{2}_{y}}{{\sigma}^{2}_{y_o}} + \frac{{\sigma}^{2}_{y}}{{\sigma}^{2}_{\eta}},
\label{Eq_CSNR_Def}
\end{equation}

\noindent where ${\sigma}^{2}_{y}$ is the variance of the ideal floating-point output, and ${\sigma}^{2}_{y_i}$ and ${\sigma}^{2}_{y_o}$ represent the variance of the output considering input and output quantization, respectively. For CSNR, the additional term ${\sigma}^{2}_{\eta}$ accounts for variations caused by analog nonidealities. Jia~\cite{Programmable_BSBP} used randomly generated weight and activation vectors, comparing the post-quantization output to the ideal full-precision output to measure the SQNR of a real device, as described by:

\begin{equation}
\begin{aligned}
\text{CSNR (SQNR)} & = \frac{\text{P}_{\text{Signal}}}{\text{P}_{\text{Noise}}} \\
& = 10 \log_{10} [\frac{\sum^{N-1}_{n=0} y^{2}_{n}}{\sum^{N-1}_{n=0} {(y_{n} - \hat{y_{n}})}^{2}}].
\end{aligned}
\label{Eq_CSNR_Measure}
\end{equation}

\noindent Here, $y_n$ is the ideal full-precision output, and $\hat{y_n}$ is the expected output from a real device. Yoshioka~\cite{CR_ACIM_JSSC} extended this approach by introducing analog Gaussian noise into each binary MAC cycle to estimate the actual CSNR of ACiM chips. While the noise terms in the denominator of Eqs.~\ref{Eq_SQNR_Def} and~\ref{Eq_CSNR_Def} can be minimized with careful circuit design, the signal term in the numerator is highly dependent on the specific DNN model and input data. Assuming a uniform distribution across the ADC’s full dynamic range can result in amplified signal variance, leading to an overestimation of CSNR. 

In this paper, we demonstrate that the output distribution in ACiM depends on factors such as the DNN model, input activations, and MAC cycle index. Consequently, CSNR may not reliably reflect inference quality in practice. Our ASiM framework offers a more representative evaluation for realistic DNN tasks and supports design space exploration grounded in accurate modeling.

\section{ASiM Framework}
\label{Chapt_Framework}

\subsection{Overview}

\begin{figure*}[t!]
\centering
\includegraphics[width=\textwidth]{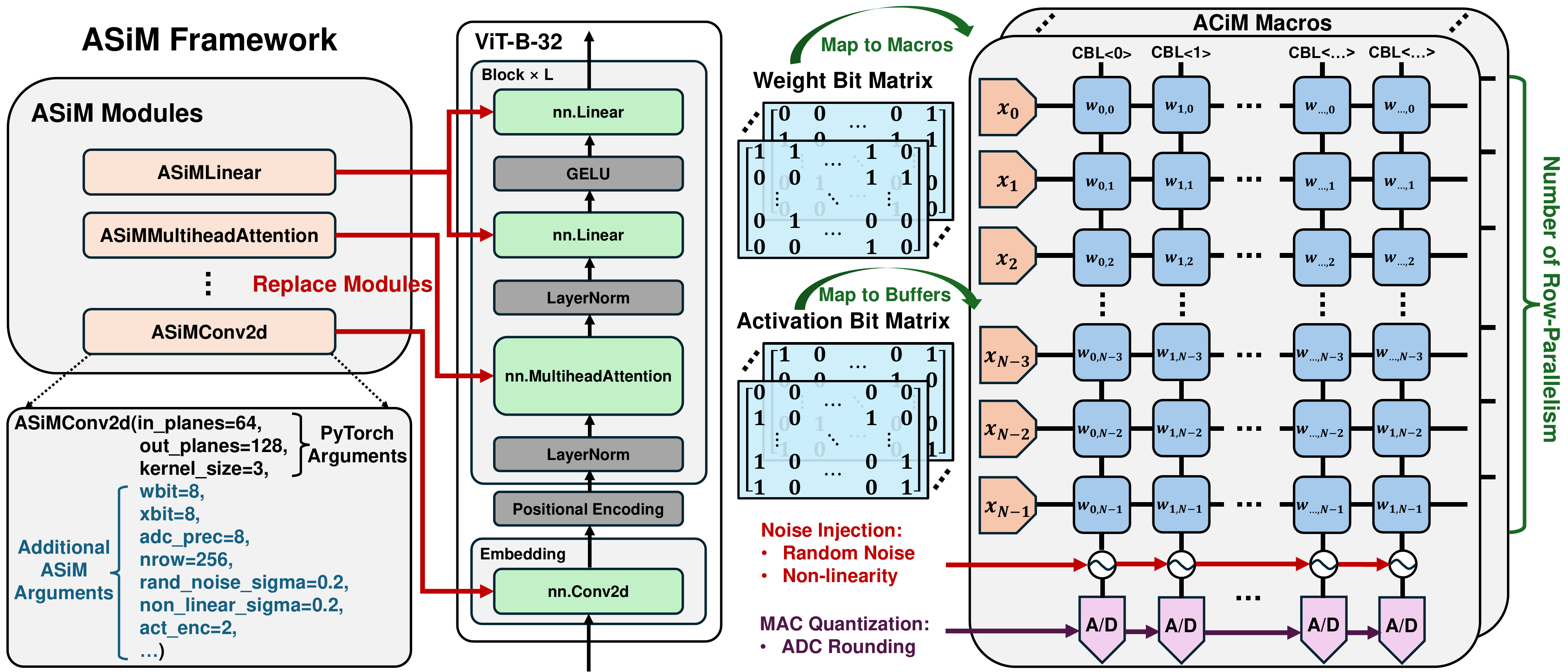}
\caption{ASiM framework overview: ASiM is a plug-and-play simulation framework tailored for charge-domain ACiM inference. Users can seamlessly replace default PyTorch modules with ASiM modules to perform inference simulations by providing additional circuit parameters. The framework automatically decomposes the weights/activations and maps it to the designated macros, following ACiM computing rules while incorporating noise and ADC effects.}
\label{Fig_ASiM}
\end{figure*}

The overview of ASiM framework is shown in Fig.~\ref{Fig_ASiM}. ASiM is implemented within PyTorch to ensure seamless integration with modern DNN workflows. It provides a plug-and-play simulation framework tailored for SRAM-based charge-domain ACiM. For MAC-heavy operations such as \texttt{CONV}, \texttt{Linear}, and \texttt{Attention}, ASiM modules can replace standard PyTorch layers, enabling bit-wise simulations with user-defined circuit parameters. 

During inference, quantized weights and activations are decomposed into binary bit tensors, split according to the predefined macro dimensions, and mapped onto the ACiM macros. The bit-wise simulation transforms the original floating-point operations into $\text{wbit} \times \text{xbit} \times \text{mapping}$ operation is simulated using native PyTorch functions, producing ideal analog voltages on CBLs. ASiM then applies analog noise and ADC effects to these voltages. By leveraging PyTorch’s CUDA acceleration, ASiM achieves both simulation fidelity and high performance. Additionally, ASiM supports bit-parallel operation, where input activations are encoded based on encoding bit width, and the framework calculates bit-parallel voltage values using pre-configured algorithms, ensuring compatibility with most latest ACiM designs.

\subsection{Simulation Algorithm and Implementation Details}

\subsubsection{Algorithm} 

A sample bit-serial simulation algorithm using the ASiM module is shown in Algorithm~\ref{Algorithm_ASiM}. The ASiM module initializes user-provided arguments that define the characteristics of the prototype ACiM circuits. Once initialized, the module quantizes all model weights and input activations, decomposes them into binary bit tensors, and splits these tensors according to the predefined macro dimensions. During the forward pass of the ASiM modules, the ideal CBL voltage for each column is computed by applying PyTorch functions such as \texttt{conv2d}, \texttt{linear}, or \texttt{matmul} on each bit-serial cycle with binary tensors. After determining the ideal voltage value for each cycle, analog noise—based on the predefined noise model—is added to these voltage values. The CBL voltage is subsequently read out by ADCs, which round these analog values to the nearest discrete level based on ADC precision, followed by bit-shifting and accumulation to yield the final output activation values, as described in Eq.~\ref{Eq_MAC}.

\begin{algorithm}[htbp]
\caption{Bit-wise Simulation of ASiM Modules}
\label{Algorithm_ASiM}
\begin{algorithmic}[1]
\REQUIRE Input activations
\ENSURE Output activations
\STATE Initialize ASiM module with the following parameters: bit width for weights and activations, ADC precision, row parallelism, analog noise intensity, and others
\STATE Quantize input activations and weights to $X$ and $W$ bits. Convert them into binary bit tensors and split them into $N$ tensors to match the dimensions of the macros
\FOR{Each mapping update: $n = 0, 1, \dots, N-1$}
\FOR{Each weight bit index: $i = 0, 1, \dots, W-1$}
\FOR{Each input bit index: $j = 0, 1, \dots, X-1$}
\STATE Perform bit-wise computation: \\ $V_{MAC} = $\texttt{nn.functional}$(w_i, x_j)$
\IF{Analog noise is included}
\STATE Generate and add noise: \\ $V_{MAC} = V_{MAC} + f_{N}(\sigma)$
\ENDIF
\STATE Apply ADC rounding function: \\ $MAC_{i,j} = f_{ADC}(V_{MAC})$
\STATE Update multi-bit output activations: \\ $\text{Output} = \text{Output} + \text{Shift}(MAC_{i,j}, i+j)$
\ENDFOR
\ENDFOR
\ENDFOR
\RETURN Output activations computed by the ACiM circuits
\end{algorithmic}
\end{algorithm}

\subsubsection{Model Quantization} 

To streamline implementation, per-tensor uniform quantization is employed as the quantization strategy for all model weights and activations in the ASiM modules. For post-ReLU activations in CNNs and post-softmax attention in ViTs, where the activation values are consistently positive, unsigned quantization is used. For all model weights and the remaining activations, signed quantization in 2's complement format is applied, where the MSB represents the sign. During the forward propagation of each layer using the ASiM module in a simulation, weight and activation tensors are quantized into integer format from floating-point values via a scaling factor. After completing the bit-wise computations, the results are converted back to floating-point format, thereby incorporating the effects of model quantization noise. Operations unrelated to ACiM, such as activation functions and batch normalization, are performed in floating-point format to exclude irrelevant influences.

\subsubsection{Mapping Strategy} 

In the ASiM configuration, we adopt the weight stationary data flow commonly used in most ACiM designs, where model weights are stored in SRAM cells, and input activations are frequently refreshed and broadcast across the macro. In the \texttt{MultiheadAttention} layers, which involve matrix multiplications (QK and AV), Queries (Q) and Values (V) are stored in SRAM cells, while Keys (K) and Attention Scores (A) are broadcast across the macro, providing an optimal configuration suitable for bit-parallel simulations.

\subsubsection{Activation Encoding}

In bit-parallel simulations, activations are converted into multi-step voltage values using the specified encoding bit width. For unsigned activations, all bits are encoded directly according to the given parameter. If the activation bit width does not evenly divide the encoding bit width, the encoding begins with the LSB, and the remaining MSB bits are encoded separately. For signed activations, the MSB (sign bit) is processed in a bit-serial manner, while the remaining bits follow the same encoding rules as unsigned activations.

\subsubsection{Analog Noise Modeling}

As detailed in Algorithm~\ref{Algorithm_ASiM}, analog noise can be added to the CBL voltage of each column to emulate real ACiM devices, enabling evaluation of its effect on inference accuracy. As ASiM is designed as a system-level simulator, it does not model or differentiate between specific low-level analog noise sources. Instead, the cumulative effect of these noise sources is abstracted as equivalent fluctuations on CBL voltage for ADC readout. Therefore, ASiM includes two simple noise models: random noise and nonlinearity, both of which can be configured with user-defined noise intensities. The combined analog noise model, incorporating these two noise types, reflects the majority of conditions found in fabricated circuits. Static offset noise, while potentially present in real circuits, can be mitigated through ADC code calibration and is therefore omitted from ASiM for simplicity. 

Random noise covers sources such as $kT/C$ noise and random comparator decision errors, which can be simply represented with a Gaussian distribution modeling an ADC input-referred noise~\cite{Noise_Modeling_SAR,Random_Decision_Error}. The distribution has a mean of zero and a standard deviation corresponding to the noise intensity, with ASiM taking this intensity as a percentage relative to the entire operational voltage range. Nonlinearity arises from device mismatches and layout-specific issues; ASiM incorporates capacitance mismatch in local memory cells and DACs, scaled by $\sqrt{N}$, where $N$ is the number of capacitors in the CBL and DAC~\cite{Cap_Non_linear}. Due to the stochastic nature of mapping DNNs to ACiM hardware, spatial noise such as mismatch and PVT variation, is expected to have a similar impact on accuracy and is thus integrated into the overall random noise and nonlinearity model in this paper. Users can also incorporate custom noise models to account for nonlinearities specific to their designs, enabling tailored inference performance analysis with ASiM. 

\subsubsection{ADC Modeling} 

ADCs convert the CBL voltage across the entire operational range into discrete steps, determined by ADC precision, as shown by:

\begin{equation}
MAC = \text{clamp} \left( \left\lfloor V_{MAC} \right\rceil, 0, 2^k - 1 \right),
\label{Eq_ADC}
\end{equation}

\noindent where $V_{MAC}$ denotes the CBL voltage before rounding, and $MAC$ represents the digital MAC value after ADC rounding. Here, $k$ is the ADC precision, and $\left\lfloor \cdot \right\rceil$ denotes the round-to-nearest operator. After rounding, the digital output is scaled back to its corresponding MAC value, followed by bit-shifting and accumulation to produce the final multi-bit output activations. 

In~\cite{Modeling_SRAM_IMC}, the ADC’s dynamic range is clipped to a narrow span to increase granularity. While this approach can theoretically reduce quantization noise, it presents two significant drawbacks: (1) analog noise becomes proportionally more significant within the narrowed range, and (2) the implementation requires complex dynamic range adjustment circuits. Thus, this method tends to degrade the ADC's Effective Number Of Bits (ENOB) in practice, potentially leading to optimistic accuracy estimations. In contrast, our ADC model captures the full range of possible CBL voltages, better reflecting real-world ACiM implementations, providing more reliable accuracy predictions while maintaining implementation simplicity.

\begin{table}[htbp]
\centering
\caption{Comparison of inference accuracy for a bit-serial ACiM macro with 256-row parallelism under a 6-bit ADC and 0.5 $\text{LSB}_{\text{rms}}$ noise, evaluated using several existing simulators with an 8b/8b ResNet-18 model on ImageNet.}
\label{Table_Preliminary_Analysis}
\begin{tabular}{c|cccc}
\toprule
\textbf{} & \textbf{CSNR} & \textbf{CIMUFAS} & \textbf{Saikia} & \textbf{NeuroSim} \\
\midrule
\textbf{ADC Model}   & SNR & Gaussian & Adaptive & Min-Max \\
\textbf{Accuracy}   & 65.44\% & 0.10\% & 40.82\% & 61.09\% \\
\bottomrule
\end{tabular}
\end{table}

\subsection{Preliminary Analysis of Existing Simulators}

We conducted a preliminary analysis using several prevailing simulation frameworks, and the predicted results are summarized in Table~\ref{Table_Preliminary_Analysis}. While each framework has its merits, differences in modeling approaches lead to divergent accuracy predictions, underscoring the importance of SRAM-specific modeling capabilities as offered by ASiM.

Our analysis identifies ADC modeling and noise representation as key factors driving the divergence in accuracy predictions. For instance, the CSNR approach~\cite{Programmable_BSBP, CR_ACIM_ISSCC, PICO-RAM} models quantization and analog noise by injecting scaled random noise into the model’s forward pass based on a predefined SNR. However, it neglects the noise amplification caused by shifting and accumulation of partial MAC results, resulting in significant discrepancies from actual SRAM-based ACiM circuit behavior. While CIMUFAS~\cite{Fast_Accurate_Sim} performs binary-level decomposition of CiM operations, it does not include an explicit ADC model and instead introduces Gaussian noise as a substitute, potentially limiting its ability to accurately reflect inference performance. Saikia’s model~\cite{Modeling_SRAM_IMC} assumes an idealized adaptive dynamic range mechanism that is not readily implementable in real ADCs, thereby underestimating the quantization error introduced by degraded ENOB in practical hardware. NeuroSim~\cite{NeuroSim,DNN+NeuroSim} quantizes ADC outputs based on the Min-Max range of MAC results, which leads to overly optimistic estimations of ADC rounding effects and the corresponding impact of analog noise. Although each framework has its strengths, these modeling differences can lead to diverging accuracy predictions. Without clear guidance, such discrepancies may inadvertently influence circuit design decisions. This highlights the need for a consistent and accurate simulation methodology to support reliable evaluation and design exploration for ACiM systems.

To address these concerns, ASiM simulates ADC behavior across its full dynamic range, accurately capturing both rounding effects and noise-induced errors based on the real MAC output distribution. Additionally, ASiM supports the latest Transformer-based models and incorporates advanced features such as bit-parallel ACiM evaluation. Lastly, ASiM’s modular design allows for seamless integration into various codebases, making it a user-friendly and adaptable tool for ACiM circuit evaluation.

\subsection{Noise-Aware Training}

Although ASiM is designed as a plug-and-play framework, directly using the default floating-point model for ACiM inference often leads to significant accuracy loss. To address this, we provide interfaces for Quantization-Aware Training (QAT) and Noise-Aware Training (NAT) for model fine-tuning~\cite{White_Paper_Quantization,Low-Cost_7T}. When set to \texttt{Train} mode and configured with the desired quantization bits for weights and activations, QAT is performed to help correct errors from model quantization. However, the model remains insufficiently robust for analog computing, as DNN models require retraining to accommodate diverse additional noise sources. To build a model that can withstand ADC quantization and analog noise simultaneously, ASiM supports NAT, enhancing model robustness as described by the equation:

\begin{equation}
\widetilde{O} = O \cdot (1 + \eta),
\label{Eq_NAT}
\end{equation}

\noindent where $O$ is the ideal output activation expected by model inference, and scaled Gaussian noise, $\eta \sim \mathcal{N}(0, \sigma^2)$, is applied directly during the floating-point forward pass to emulate the noisy outputs generated by ACiM circuits. By fine-tuning the DNN model with various training noise intensities $\sigma$, we select a model that maintains acceptable baseline accuracy while maximizing noise tolerance for ACiM applications.

\begin{figure}[htbp]
    \centering
    \includegraphics[width=\columnwidth]{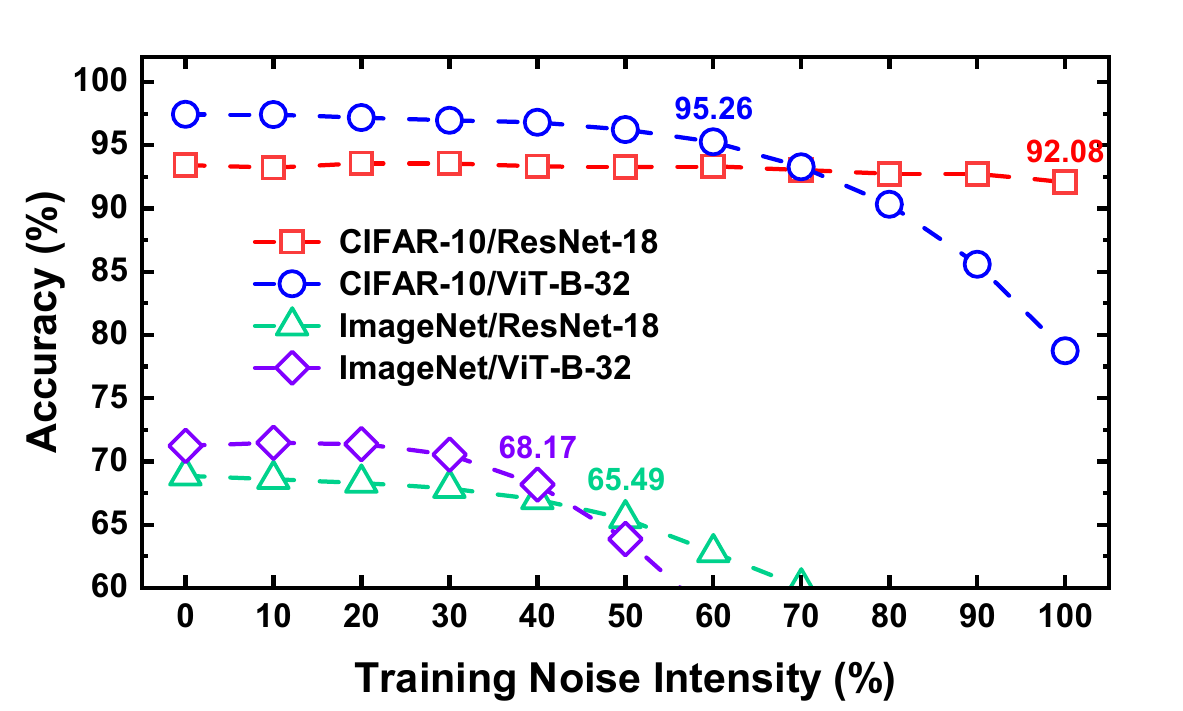}
    \caption{Noise-Aware Training: accuracy of 8b/8b baseline models across different training noise intensities. ResNet models demonstrate greater robustness compared to ViT, while ImageNet proves more challenging than CIFAR-10.}
    \label{Fig_NAT}
\end{figure}

We conducted NAT on ResNet-18 and ViT-B-32 using the CIFAR-10 and ImageNet datasets, respectively, with the baseline model accuracy under different training noise intensities value of $\sigma$ shown in Fig.~\ref{Fig_NAT}. As observed in many previous studies, ResNet models tend to be easier to fine-tune compared to ViT models, and the ImageNet task is generally more challenging than CIFAR-10. Consequently, we selected models optimized for ACiM applications in the following analysis, with a 2.0\% accuracy drop for the CIFAR-10 task and around a 3.0\% accuracy drop for the ImageNet task.

\section{Performance Analysis of ACiM Inference}
\label{Chapt_Analysis}

\subsection{Re-consider CSNR with Actual Data Distributions}

\begin{figure}[htbp]
    \centering
    \subfloat[]{%
        \includegraphics[width=0.98\linewidth]{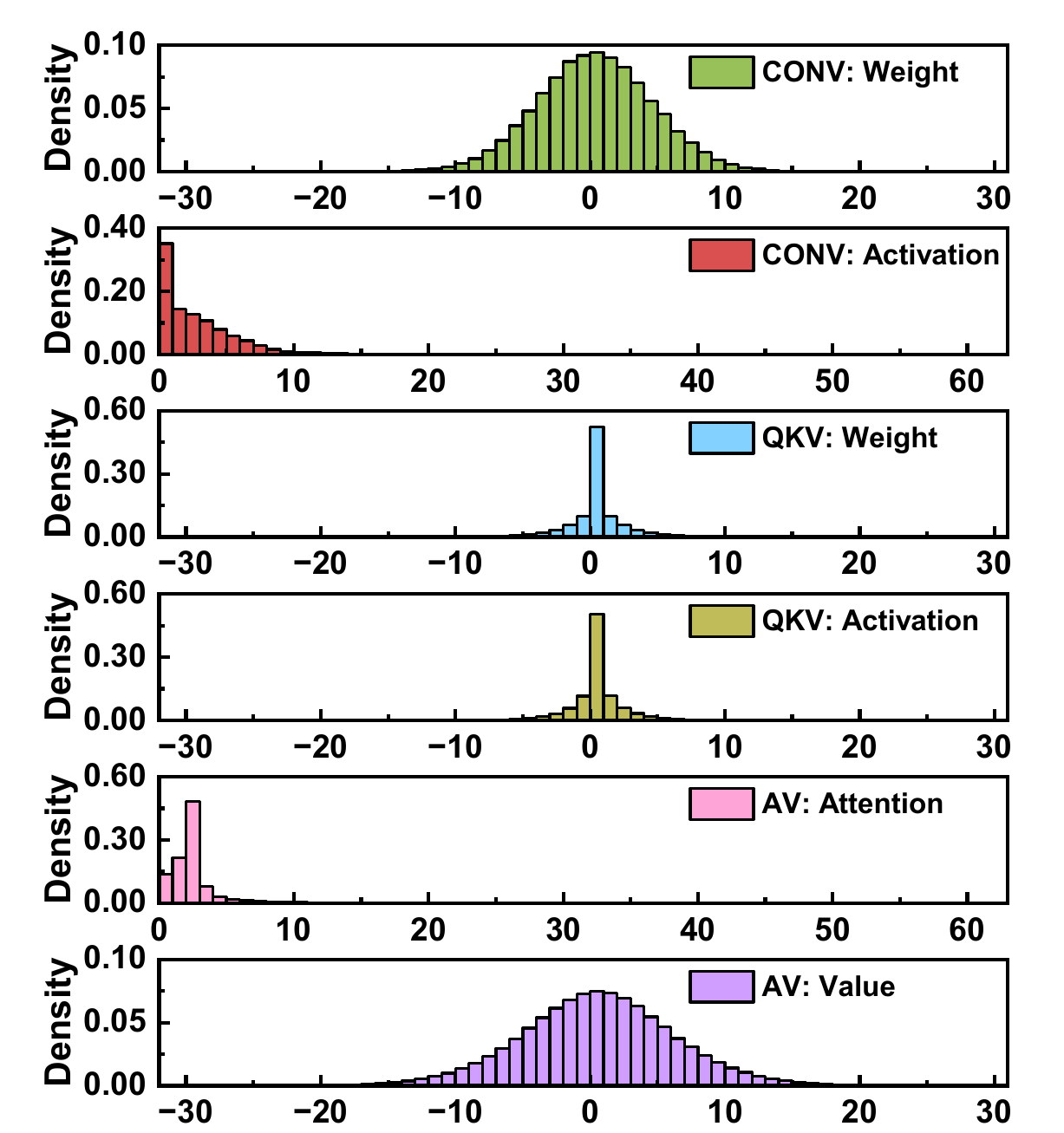}
        \label{Fig_Input_Dist_a}
    }\\

    \subfloat[]{%
        \includegraphics[width=0.98\linewidth]{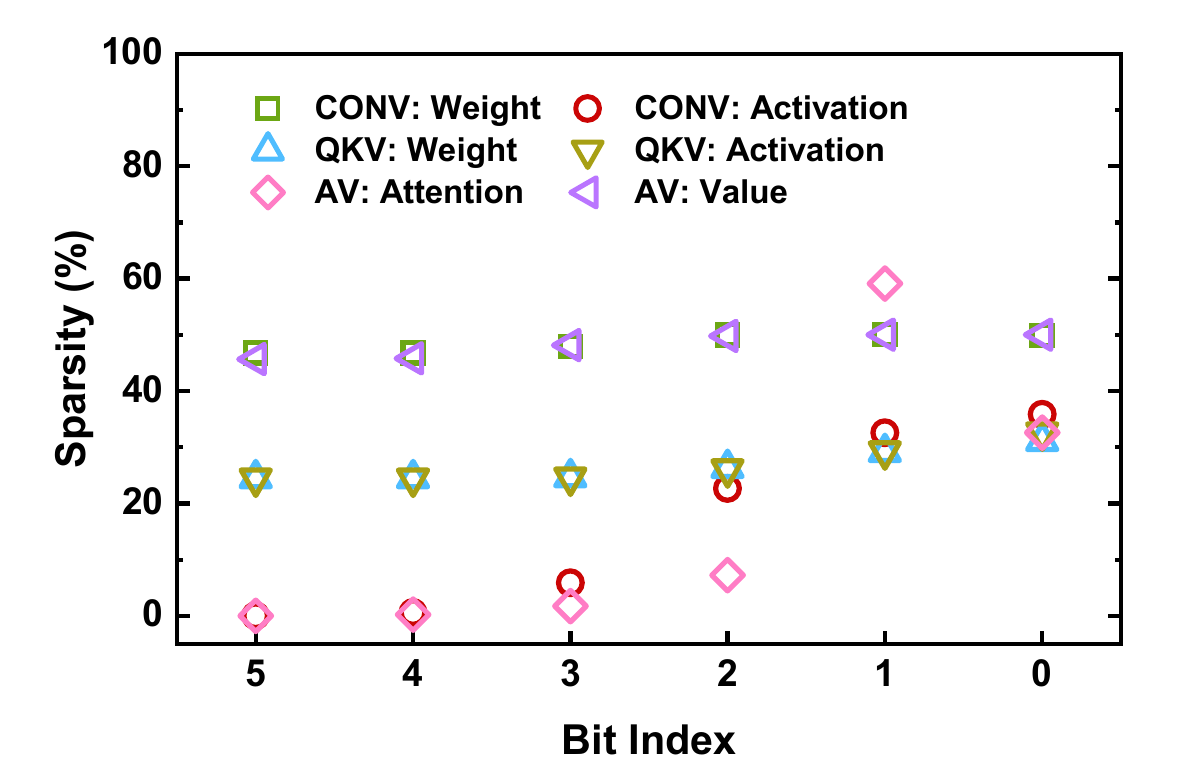}
        \label{Fig_Input_Dist_b}
    }%
    
    \caption{Quantized DNN model input distribution analysis. (a) Data distribution of representative DNN layers after 6b/6b quantization. (b) Bit-level sparsity for each cluster in (a), measured per bit index, where sparsity represents the ratio of 1's at each bit position.}
    \label{Fig_Input_Dist}
\end{figure}

To establish a foundation for evaluating how different design factors impact inference accuracy, we first analyzed the CSNR under realistic neural network data distributions. Our analysis focuses on ResNet-18 and ViT-B-32 architectures using the CIFAR-10 dataset, with both models configured for 6-bit weight and 6-bit activation quantization. The distributions of several common layers processed by ACiM are shown in Fig.~\ref{Fig_Input_Dist}(a), with their bit-level sparsity for each bit index displayed in Fig.~\ref{Fig_Input_Dist}(b). In convolutional layers, we observed distinct patterns between weights and activations:
\begin{itemize}[leftmargin=*]
\item Activations show a characteristic of one-sided distribution due to \texttt{ReLU} operations that zero out negative values.
\item Activation bit-level sparsity remains consistently below 40\%, with MSBs being particularly sparse.
\item Weights exhibit a normal distribution, resulting in approximately 50\% bit-level sparsity across all bit positions.
\end{itemize}

\noindent The ViT architecture presents more complex distribution patterns and distinctive sparsity distribution. In the linear layers:
\begin{itemize}[leftmargin=*]
\item QKV projection layers show significant cross-channel variations due to \texttt{LayerNorm} operations~\cite{Challenges_Transformer_Quantization}.
\item Both weights and activations display two-sided distributions with numerous small values quantized to zero in per-tensor quantization configuration.
\item By these reasons, bit-level sparsity in QKV projections consistently ranges between 20\% and 30\% for both weights and activations.
\end{itemize}

\noindent The attention computation demonstrates particularly diverse characteristics:
\begin{itemize}[leftmargin=*]
\item Value (V) matrices follow a normal distribution with approximately 50\% bit-level sparsity.
\item In most cases, attention scores (A) exhibit a long-tail distribution characterized by predominantly small values with occasional large outliers, resulting in bit-level sparsity varies dramatically from 0\% to 60\%, with MSBs showing exceptionally high sparsity.
\end{itemize}

\begin{figure*}[t!]
\centering
\includegraphics[width=\textwidth]{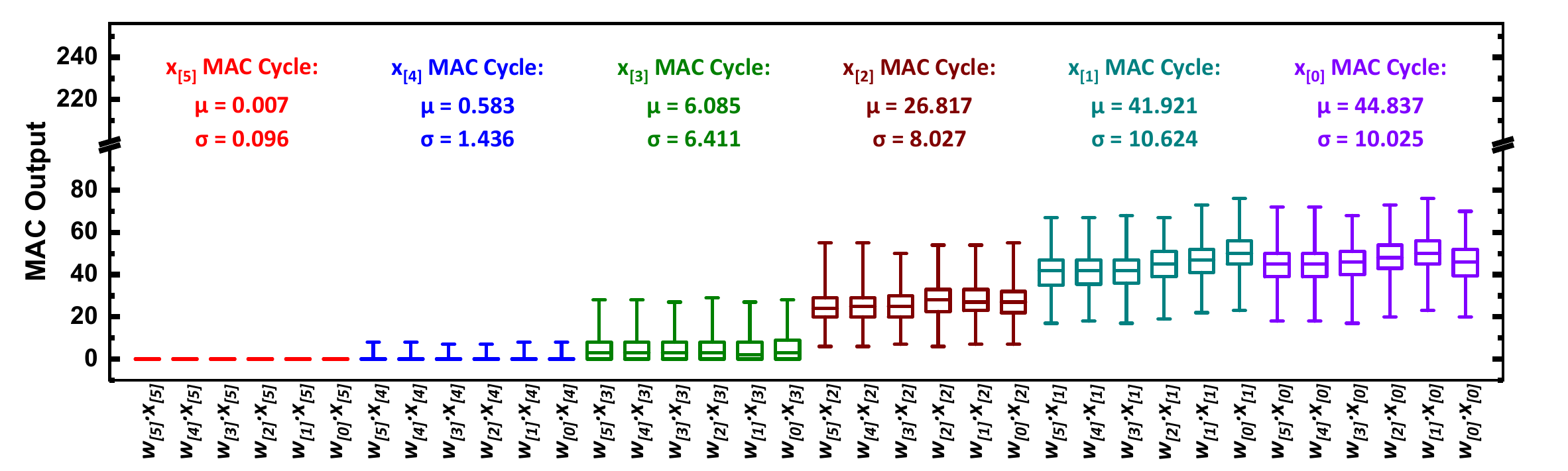}
\caption{Distribution of MAC outputs in the convolutional layer of a 6b/6b CNN for each bit-serial cycle, using an ACiM macro with row parallelism of 256 and 8-bit ADC. The MAC outputs associated with the MSBs of the activation are consistently low, while outputs increase as the activation bit index moves toward the LSBs. Still, the maximum signal span across all cycles remains only 50 within the full ADC dynamic range of 256.}
\label{Fig_MAC_Dist}
\end{figure*}

The MAC output for each binary cycle is statistically proportional to the bit-level sparsity of weights and activations \cite{PACiM}, implying that the expected CBL voltage depends on the cycle indices and is generally constrained to a narrow range, rather than utilizing the full ADC dynamic range. Fig.~\ref{Fig_MAC_Dist} illustrates an example of bit-serial MAC outputs for \texttt{CONV} layer inputs from Fig.~\ref{Fig_Input_Dist}, using an ACiM macro with 256-row parallelism and an 8-bit ADC. The MAC output corresponding to the MSB of activations is nearly zero, with a very low probability of being non-zero due to the minimal bit-level sparsity in the activation. As the activation bit index shifts toward the LSBs, the MAC output increases, driven by rising bit-level sparsity. However, the maximum MAC output span in the \texttt{CONV} layer remains only 50 within the full ADC dynamic range of 256. This suggests that the signal variance in Eq.~\ref{Eq_CSNR_Def} is not fixed but is data-dependent and remains confined to a limited range. The near-zero signal span in the MSB cycles leads to a signal variance close to zero, making the output highly susceptible to noise. Conventional estimates, assuming uniform distribution of weights and inputs \cite{Programmable_BSBP,CR_ACIM_JSSC}, predict a signal variance of $\text{DR}_{\text{ADC}}^{2}/12$, which is significantly higher than the actual variance, resulting in optimistic CSNR estimates. This observation holds for all types of DNN layers, as indicated by the sparsity in Fig.~\ref{Fig_Input_Dist}(b), motivating us to comprehensively verify how each design factor influences inference performance in real DNN applications.

\subsection{ADC Quantization Noise}

We begin our exploration of inference accuracy by examining the quantization noise introduced by ADCs in ACiM circuits. Unlike Digital Compute-in-Memory (DCiM), where inference accuracy is influenced solely by model quantization, the ADC rounding effect in ACiM circuits introduces additional quantization errors to each MAC output.

\textbf{Boundary Precision.} We define ADC boundary precision as the precision that corresponds exactly to the row-parallelism of the ACiM. For example, in a conventional bit-serial ACiM with a row-parallelism dimension of $2^k$, the boundary precision is $k$, which would ideally translate the CBL voltage into digital code. Using an ADC with precision lower than $k$ introduces quantization errors that could degrade inference accuracy, while ADC precision higher than $k$ is redundant and adds unnecessary area and energy overhead. To provide a quantitative example of how reducing ADC precision affects performance, we consider an ACiM macro configured with a row parallelism of 256 and 4-bit activation encoding. In this configuration, each local capacitor can represent 16 distinct states (from 0 to 15), and a lossless readout must span the entire computational range from 0 to 3,840 to accurately capture all possible MAC outputs. As a result, achieving error-free readout requires an ADC boundary precision of 12 bits to avoid any rounding-induced errors.

\textbf{Trade-offs and Design Policy.} While increasing row-parallelism dimension improves ACiM throughput, it also requires higher ADC precision to maintain the same computational accuracy. In this paper, we selected a row parallelism of 256 as it provides an optimal balance between throughput and tunable ADC precision. An optimized design carefully selects ADC precision slightly below the boundary precision, balancing minor accuracy loss with significant efficiency gains.

\textbf{Experiments.} We performed inference simulations on the CIFAR-10 and ImageNet datasets using the ResNet-18 CNN model and the ViT-B-32 Transformer model, respectively. Both models are trained with 8b/8b weight/activation QAT, followed by fine-tuning with appropriate noise intensities through NAT. We analyzed the accuracy loss due to quantization noise for both bit-serial and bit-parallel scheme to identify the optimal ACiM settings for each model and dataset.

\subsubsection{Experimental Analysis on Bit-serial Scheme}

\begin{figure}[htbp]
    \centering
    \includegraphics[width=\columnwidth]{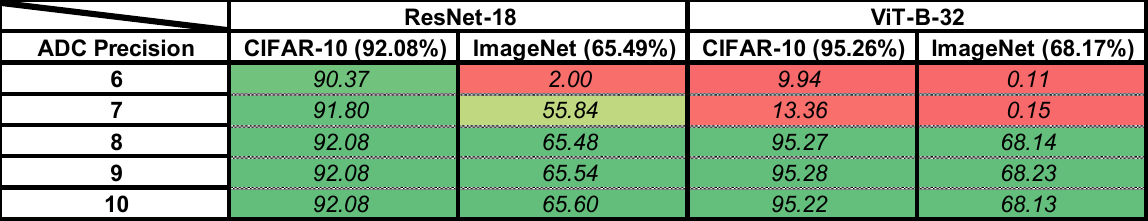}
    \caption{Shmoo plot of inference accuracy under varying ADC precision for a bit-serial ACiM macro with a row-parallelism of 256, evaluated using an 8b/8b DNN model. The accuracy of the digital counterpart is provided alongside each model’s name for reference.}
    \label{Fig_Accuracy_BS}
\end{figure}

To assess the minimum ADC bit precision necessary for bit-serial ACiM, we performed simulations across various ADC precisions on a macro with row parallelism of 256, with results shown in Fig.~\ref{Fig_Accuracy_BS}. For ResNet-18 on CIFAR-10, ADC precision can be lowered by up to 2 bits below the boundary with minimal accuracy loss. However, on the more challenging ImageNet dataset, lowering ADC by even 1 bit below the boundary precision leads to an accuracy loss of approximately 10\% compared to the baseline, indicating potential issues in practical applications. In contrast, any reduction in ADC precision below the boundary causes inference accuracy to approach random guessing in ViT-B-32. Fig.~\ref{Fig_Accuracy_BS} thus confirms that Transformer models are generally more sensitive to ADC precision than CNNs, and ImageNet is indeed a more complex task than CIFAR-10, in line with design intuition. Our findings suggest that: 
\begin{itemize}[leftmargin=*]
    \item Only CNN-based models targeting CIFAR-10-level tasks can afford ADC precision reductions below boundary precision.
    \item More complex models and tasks must retain ADC precision at or above boundary precision.
\end{itemize}

\subsubsection{Experimental Analysis on Bit-parallel Scheme}

We proceed to analyze the accuracy degradation caused by quantization noise in bit-parallel ACiM. Simulations were performed on a macro with row parallelism of 256, increasing ADC precision from 8 bits to determine the best configuration for each DNN model and dataset in bit-parallel scheme. Encoding bit widths were chosen as 2 and 4, and results were compared with the bit-serial configuration. Activations were quantized to 9 bits for signed data in 2’s complement format, with the sign bit processed in bit-serial mode while the remaining bits were evenly encoded within 2-bit and 4-bit widths to allow a fair comparison with unsigned activations. Notably, each additional encoding bit requires a one-bit increase in ADC boundary precision in bit-parallel scheme, as higher ADC resolution is needed to eliminate quantization errors from encoding. 

\begin{figure*}[ht]
\centering
\includegraphics[width=\textwidth]{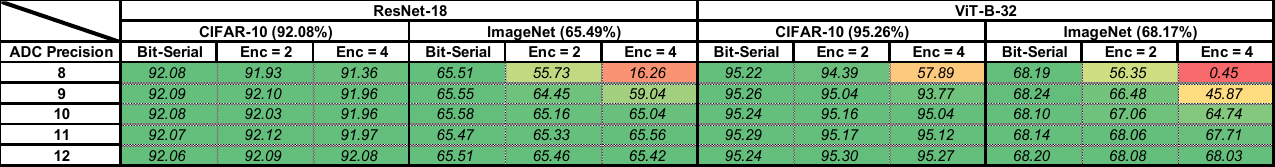}
\caption{Shmoo plot of inference accuracy under varying ADC precision for a bit-parallel ACiM macro with a row-parallelism of 256, evaluated using an 8b/8b DNN model. Results for encoding bit widths of 2 and 4 are shown and compared with those of the bit-serial counterparts. The accuracy of the digital counterpart is provided alongside each model’s name for reference.}
\label{Fig_Accuracy_BP}
\end{figure*}

As illustrated in Fig.~\ref{Fig_Accuracy_BP}, ResNet-18 on CIFAR-10 can encode activations to 4 bits while retaining 8-bit ADC precision with minimal accuracy loss. For ViT-B-32, 2-bit activation encoding preserves ADC precision at 8 bits, while 4-bit encoding requires a 1–2 bit increase in ADC precision for comparable inference performance. On the more challenging ImageNet task with ResNet-18, 2-bit and 4-bit encoding require 1 and 2 bits of additional ADC precision to reach accuracy equivalent to bit-serial scheme, respectively. For ViT-B-32, achieving lossless inference requires a 2 and 3 bits increase in ADC precision for 2-bit and 4-bit encoding, respectively. 

We evaluate the trade-off between energy savings and accuracy loss using the design parameter of the prototype chip in \cite{CR_ACIM_ISSCC}, with results shown in Fig.~\ref{Fig_BP_Cost}. For our prototype, the energy required to increase ADC precision by one bit remains below a 30\% rise up to 11 bits, while encoding an additional bit for activations reduces ACiM cycles by half, yielding significant energy gains. Thus, encoding activations with 4 bits results in nearly a fourfold reduction in energy per classification across most models and datasets, with minimal impact on inference accuracy. Even in the challenging combination of ViT-B-32 and ImageNet, 4-bit activation encoding reduces energy consumption by roughly 40\%, indicating that combining higher ADC precision with more compressed activation bits per ACiM cycle achieves both energy efficiency and high inference accuracy. Our experimental analysis stresses on the following crucial points:

\begin{figure}[htbp]
    \centering
    \includegraphics[width=\columnwidth]{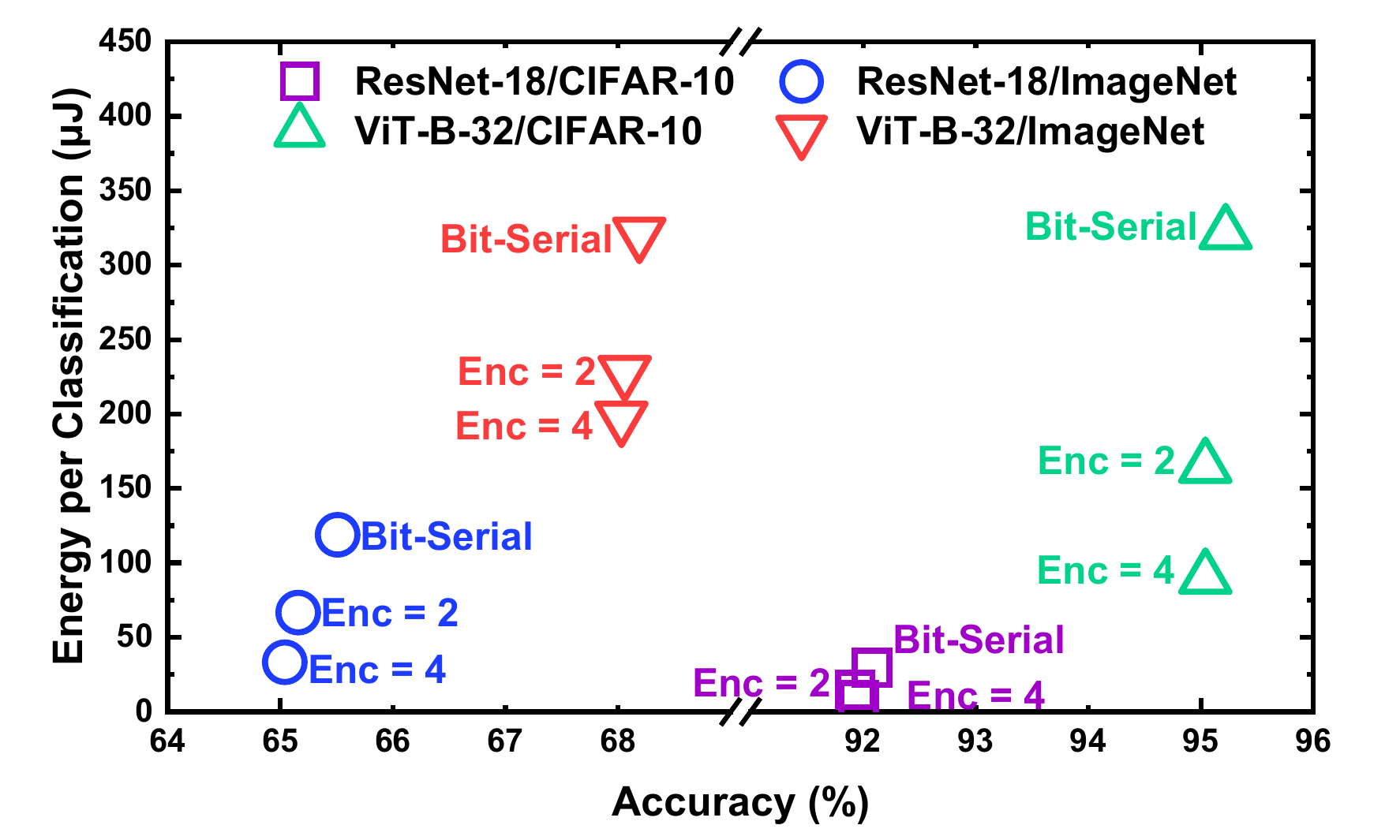}
    \caption{Energy cost and corresponding inference accuracy for bit-serial and bit-parallel ACiM configurations across various DNN models and datasets. Compressing more bits within a single cycle yields significant energy savings.}
    \label{Fig_BP_Cost}
\end{figure}

\begin{itemize}[leftmargin=*]
    \item Quantization error grows with increasing encoding width but decreases with higher ADC precision. However, pushing ADC to the boundary precision is not always necessary in bit-parallel ACiM.
    \item Higher activation encoding bit compresses more information into each LSB, giving the bit-parallel MAC a higher weight in the output activations when converted back to floating-point values. Since ADC quantization error consistently remains under 0.5 LSB, configuring bit-parallel scheme with appropriate activation encoding and ADC precision can create an optimal combination that maintains noise intensity to manageable levels in Fig.~\ref{Fig_NAT}, while enabling energy-efficient DNN inference.
\end{itemize}

\subsection{Analog Noise}
\label{Sec_Analog_Noise}
Although our previous analysis demonstrated that ADC quantization error minimally impacts DNN inference accuracy, making bit-parallel scheme viable for ACiM efficiency improvements, this assumption is established on ideal noise-free conditions. Real ACiM accelerators face various additional noise sources that require careful consideration. To assess this, we incorporated random noise and nonlinearity into simulations, analyzing their effects on inference accuracy. As detailed in Section~\ref{Chapt_Framework}, random noise includes $kT/C$ noise and comparator decision errors, while nonlinearity accounts for layout-specific issues and device mismatch, analyzed here based on capacitor mismatch in the CBL and DAC.

\textbf{Noise Definition.} In the following analysis, analog noise is modeled as a Gaussian distribution, with its intensity characterized by the following two metrics: \textbf{(1)} $V_{pp} (\%)$: This metric provides a direct representation of noise standard deviation as a percentage of the full dynamic range in the CBL. For instance, a noise level of 0.1 in $V_{pp} (\%)$ corresponds to a standard deviation of $\sigma = 1$ mV when $V_{pp} = 1$ V. \textbf{(2)} $\text{LSB}_{\text{rms}}$: Since most ACiM studies report noise levels using the Root-Mean-Square (RMS) value relative to the ADC’s LSB~\cite{C3SRAM,MACC-SRAM,In-SRAM-ADC_ACIM,PIMCA,PICO-RAM,Programmable_BSBP,CR_ACIM_ISSCC}, we also provide $\text{LSB}_{\text{rms}}$ as a reference. This metric offers a more intuitive understanding of noise characteristics in relation to the column ADC, making it particularly useful for ACiM designers.

\begin{figure*}[ht]
    \centering
    \subfloat[]{%
        \includegraphics[width=0.99\textwidth]{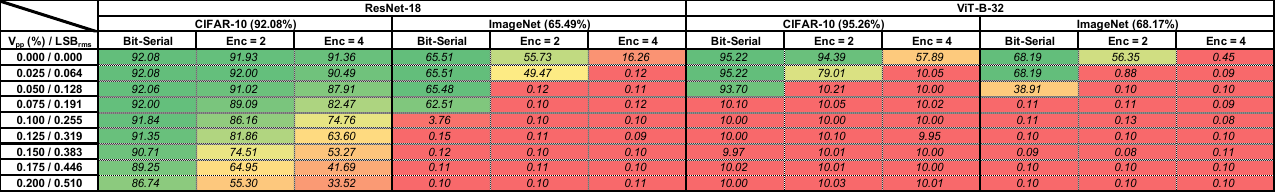}
        \label{Fig_Accuracy_Noise_a}
    }\\

    \subfloat[]{%
        \includegraphics[width=0.99\textwidth]{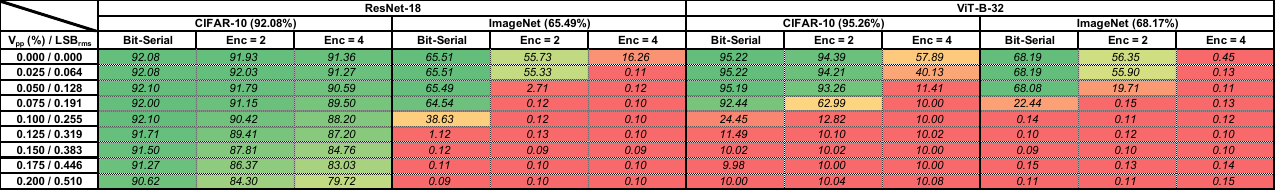}
        \label{Fig_Accuracy_Noise_b}
    }%
 
    \caption{Shmoo plot of inference accuracy across different noise intensities for an ACiM macro with a row-parallelism of 256 using an 8-bit ADC, evaluated by an 8b/8b DNN model. ResNet-18 is trained with 100\% and 50\% , and ViT-B-32 is trained with 60\% and 40\% training noise intensities for CIFAR-10 and ImageNet, respectively. Two noise models are assessed separately: (a) Random noise and (b) Nonlinearity. The accuracy of the digital counterpart is provided alongside each model’s name for reference.}
    \label{Fig_Accuracy_Noise}
\end{figure*}

Fig.~\ref{Fig_Accuracy_Noise}(a) shows a shmoo plot of accuracy for a 256-row-parallel ACiM macro with an 8-bit ADC at various random noise intensities, revealing that analog noise can heavily impact ACiM circuit inference. Unlike quantization, which limits maximum error to 0.5 LSB of the ADC, analog noise can produce readout errors on CBL voltage exceeding 1 LSB, introducing greater inaccuracies into final output activations. This effect is amplified in bit-parallel scheme, where each LSB error has an even larger effect on the final activation.
Our findings indicate task-specific noise tolerance thresholds: 

\textbf{Simple Tasks (CIFAR-10 \& CNN).} In these simple tasks, noise intensity should be kept below 0.15 $V_{pp} (\%)$ ($\sim$ 0.4 $\text{LSB}_{\text{rms}}$) to maintain inference accuracy above 90\%. This requirement tightens to 0.025 $V_{pp} (\%)$ ($\sim$ 0.1 $\text{LSB}_{\text{rms}}$) for 4-bit activation encoding in bit-parallel scheme, setting a stringent standard for real-world use. 

\textbf{Complex Tasks (ViT or ImageNet).} For complex tasks, accuracy drops sharply with increased noise intensity, making the requirements even stricter. In such cases, with an operational voltage $V_{pp}$ of 1 V for bit-serial scheme, noise intensity needs to be controlled below 0.5 mV to enable functional classification in silicon. 

In the nonlinearity analysis shown in Fig.~\ref{Fig_Accuracy_Noise}(b), a similar trend appears, with the transition boundary shifting slightly downward as nonlinearity diminishes with increased ADC output code. However, as seen in Fig.~\ref{Fig_MAC_Dist}, the MAC output is concentrated in lower-valued regions, making nonlinearity effects still significant.

\begin{figure}[htbp]
    \centering
    \includegraphics[width=\columnwidth]{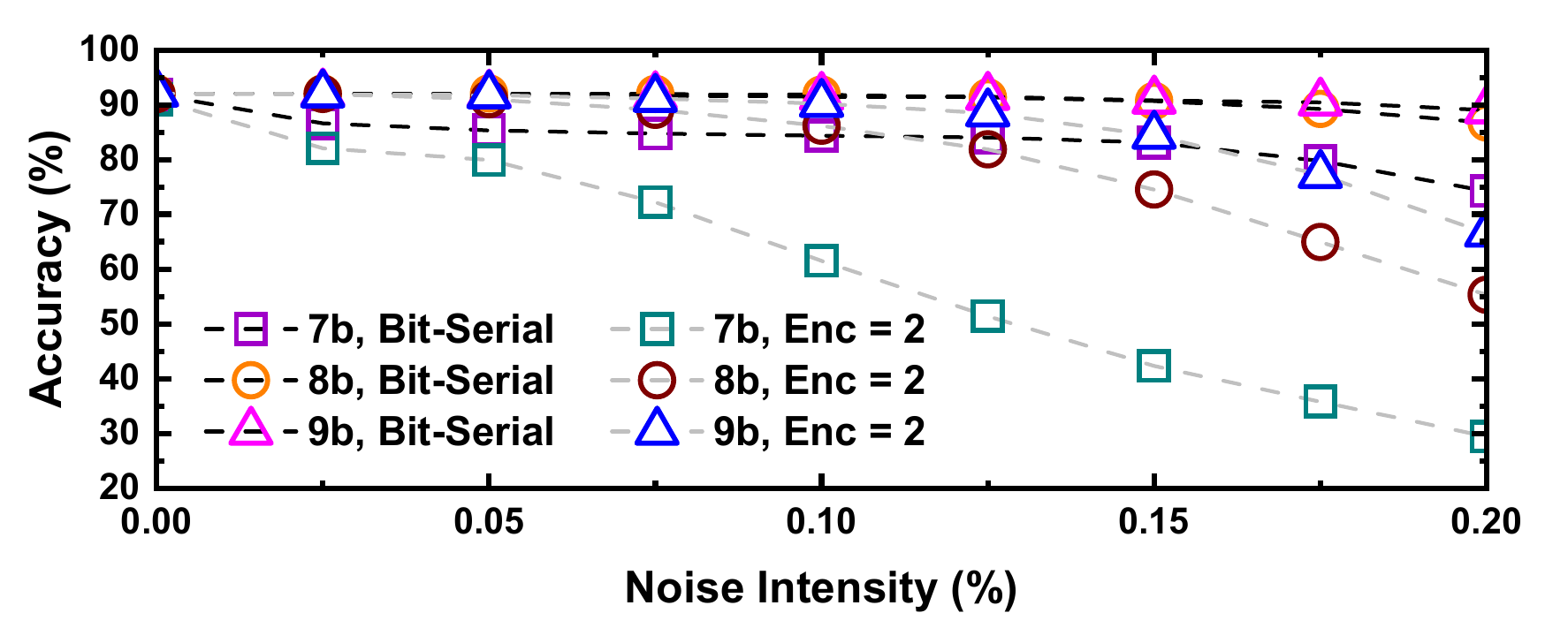}
    \caption{Accuracy degradation under different noise intensities for bit-serial and bit-parallel ACiM with varying ADC precision levels on ResNet-18/CIFAR-10. Increasing ADC precision mitigates the effects of noise.}
    \label{Fig_Noise_ADC_Dep}
\end{figure}

At a specific noise level, lower ADC precision can round out smaller errors in the CBL voltage, but any readout errors that do occur may result in greater inaccuracies in the output activation. Conversely, higher ADC precision increases the likelihood of errors but minimizes their overall impact on inference quality. This leads to a key design question: how to choose ADC precision to optimize inference accuracy. Leveraging ASiM, we evaluated accuracy degradation across varying noise intensities and ADC precision for both bit-serial and bit-parallel ACiM. The results in Fig.~\ref{Fig_Noise_ADC_Dep} indicate that as ADC precision increases, the rate of accuracy degradation decreases, highlighting the benefit of higher ADC resolution for robust inference.

\begin{figure}[htbp]
    \centering
    \includegraphics[width=\columnwidth]{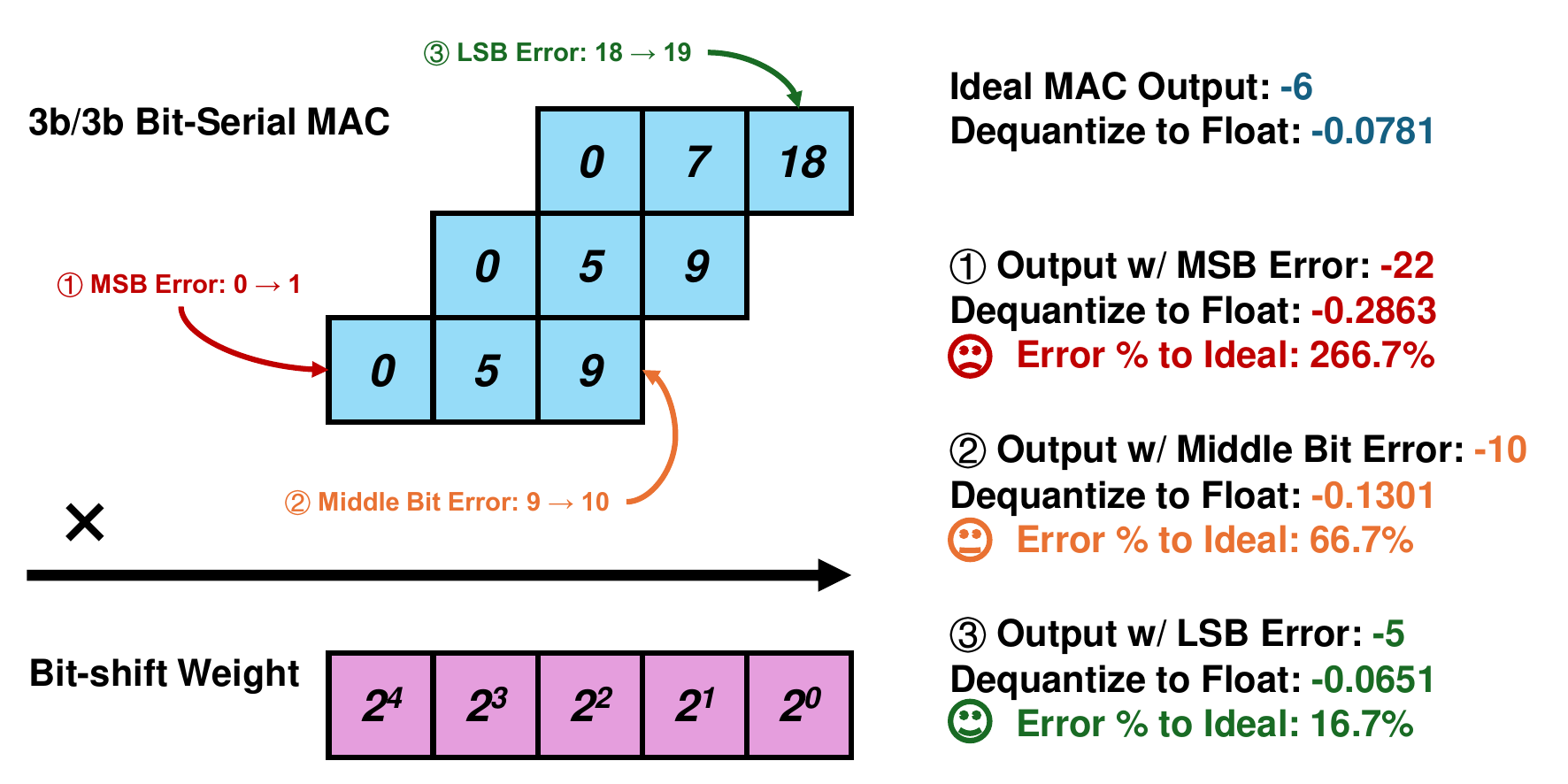}
    \caption{Real-world example of bit-serial MAC in 3b/3b ResNet-18 on CIFAR-10 with 256-row-parallel macro by 8-bit ADC. A 1-bit error has catastrophic effects in MSB cycles but is more tolerable in LSB cycles.}
    \label{Fig_Error_Demo}
\end{figure}

To validate our findings on the impact of analog noise, we conducted a detailed examination of the ACiM inference process using a 3b/3b ResNet-18 on CIFAR-10 (Fig.~\ref{Fig_Error_Demo}). Here, blue squares represent binary cycles with their respective bit indices, while purple squares indicate their bit-shift weights. In line with Fig.~\ref{Fig_MAC_Dist}, binary MAC values are relatively small within the 256 range. A 1-bit error introduced by ADC readout during MSB cycles can greatly alter the final output activation due to the amplification effect of bit-shift weight. The impact of noise manifests through two primary pathways:

\begin{itemize}[leftmargin=*]
    \item First, an 1-bit error introduced by ADC readout during MSB cycles can dramatically alter the final output activation due to the amplification effect of bit-shift weight. These large activation errors propagates through layers and severely compromise the classifier's decision accuracy.
    \item Second, these errors generate outliers in the output layers, expanding the quantization scale in subsequent layers and further increasing model quantization error.
\end{itemize}

\noindent Consequently, inference accuracy degrades sharply when errors occur during MSB cycles, and unfortunately, ACiM cannot control in which cycle the errors arise. However, if the error occurs in lower bit cycles, the inaccuracy on output activation can be reduced to less than 70\%, staying within the range where NAT can improve robustness and mitigate errors, as shown in Fig.~\ref{Fig_NAT}.

\subsection{Simulation to Silicon Correlation}

\begin{figure}[htbp]
    \centering
    \includegraphics[width=\columnwidth]{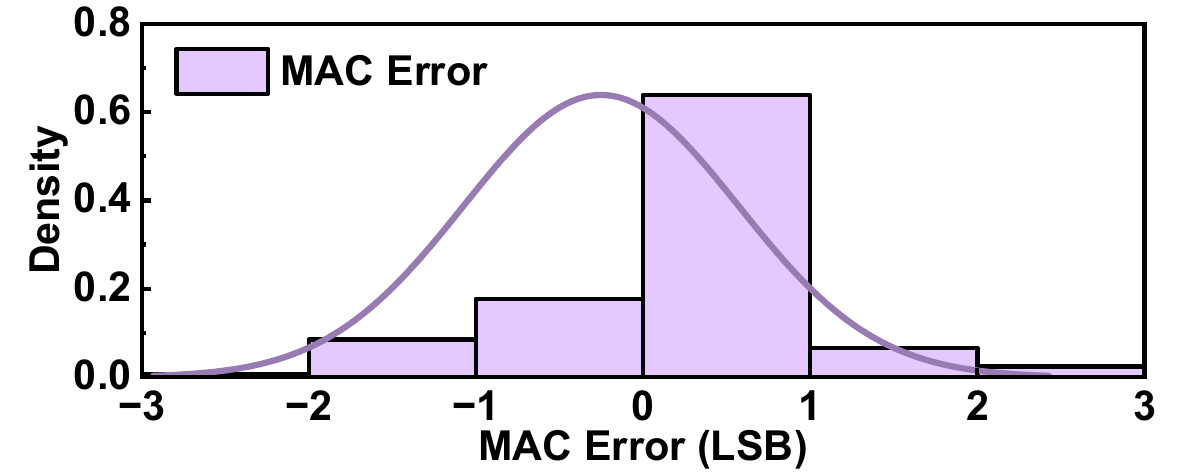}
    \caption{Error distribution in ADC output recorded from an ACiM prototype chip~\cite{CR_ACIM_ISSCC} over 30,000 trials.}
    \label{Fig_Silicon_Error}
\end{figure}

\textbf{ADC Error Distribution.} To verify that the noise model accurately reflects real ADC error characteristics, we generate random weight and input activation vectors, perform MAC operations on our prototype ACiM chip~\cite{CR_ACIM_ISSCC}, and measure the resulting voltage using an ADC, comparing it with the ideal output. This process is repeated 30,000 times, and the recorded error distribution is shown in Fig.~\ref{Fig_Silicon_Error}. As depicted, the error exhibits a bell-shaped pattern, where most ADC readings are correct, but occasional errors occur in the range of 1 to 3 LSBs. This distribution aligns well with our noise modeling assumption, validating the Gaussian noise model used in ASiM as an accurate representation of the noise encountered in ACiM chips.

\textbf{Silicon Validation.} To validate the accuracy of ASiM in predicting inference results for real chips, we compared its ResNet-18 accuracy predictions with the silicon validation results of our prototype chip reported in CR-CIM~\cite{CR_ACIM_ISSCC}. In the silicon test, CR-CIM processed the compute-intensive \texttt{CONV} and \texttt{Linear} layers of an 8b/8b ResNet-18 on-chip, while the remaining layers were handled off-chip, replicating the same setup used in ASiM simulations. The measured $\text{LSB}_{\text{rms}}$ noise was equally divided and allocated into random noise and nonlinearity. ASiM predicted an accuracy of 91.84\%, which closely corresponded to the 91.70\% measured in CR-CIM. Furthermore, we configured our prototype chip with varying ADC precision levels, and ASiM successfully identified the inflection point at which inference accuracy begins to degrade. The predicted trend closely matches our silicon measurements, supporting the validity of ASiM in evaluating inference accuracy.

\section{Towards Accurate ACiM Inference}
\label{Chapt_Accurate_Inference}

In prior sections, we performed a detailed analysis of inference accuracy using ACiM circuits and explained the mechanism how these noise degrades inference, highlighting that analog noise is a fundamental barrier to achieving high accuracy. This challenge is difficult to overcome through circuit design alone, as achieving a noise-free analog design is extremely difficult, especially in complex tasks. However, we can still optimize the application of ACiM circuits to mitigate inference accuracy concerns while leveraging their energy efficiency benefits. This section introduces and analyzes two simple and practical methods for improving inference accuracy: HCiM and majority voting. These two approaches employ a trade-off strategy to enhance inference accuracy. Specifically, HCiM sacrifices area and energy efficiency, while majority voting compromises both energy efficiency and throughput in exchange for improved accuracy.

\subsection{Hybrid CiM}

\begin{figure}[htbp]
    \centering
    \subfloat[]{%
        \includegraphics[width=0.98\linewidth]{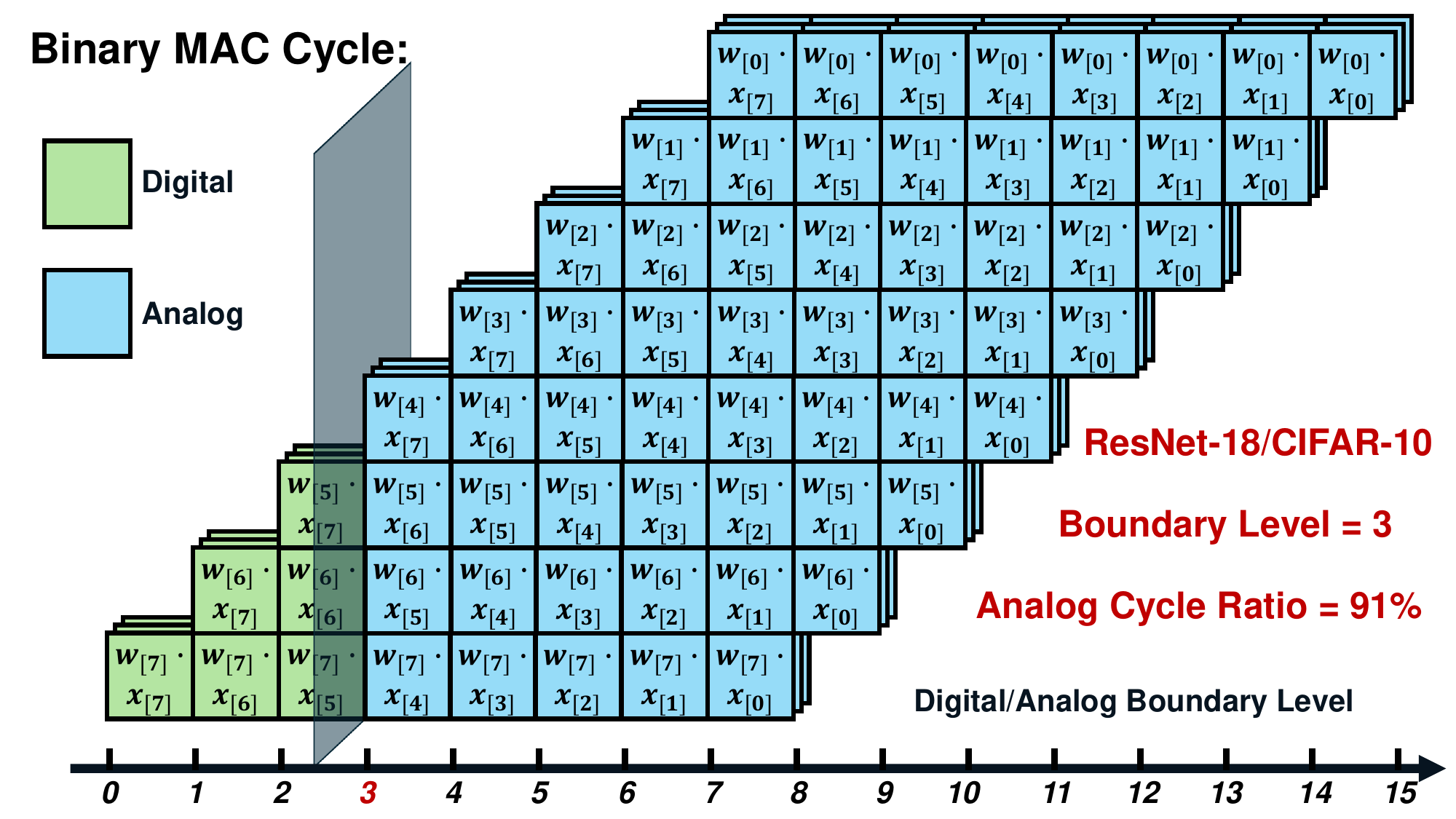}
        \label{Fig_HCiM_a}
    }\\

    \subfloat[]{%
        \includegraphics[width=0.98\linewidth]{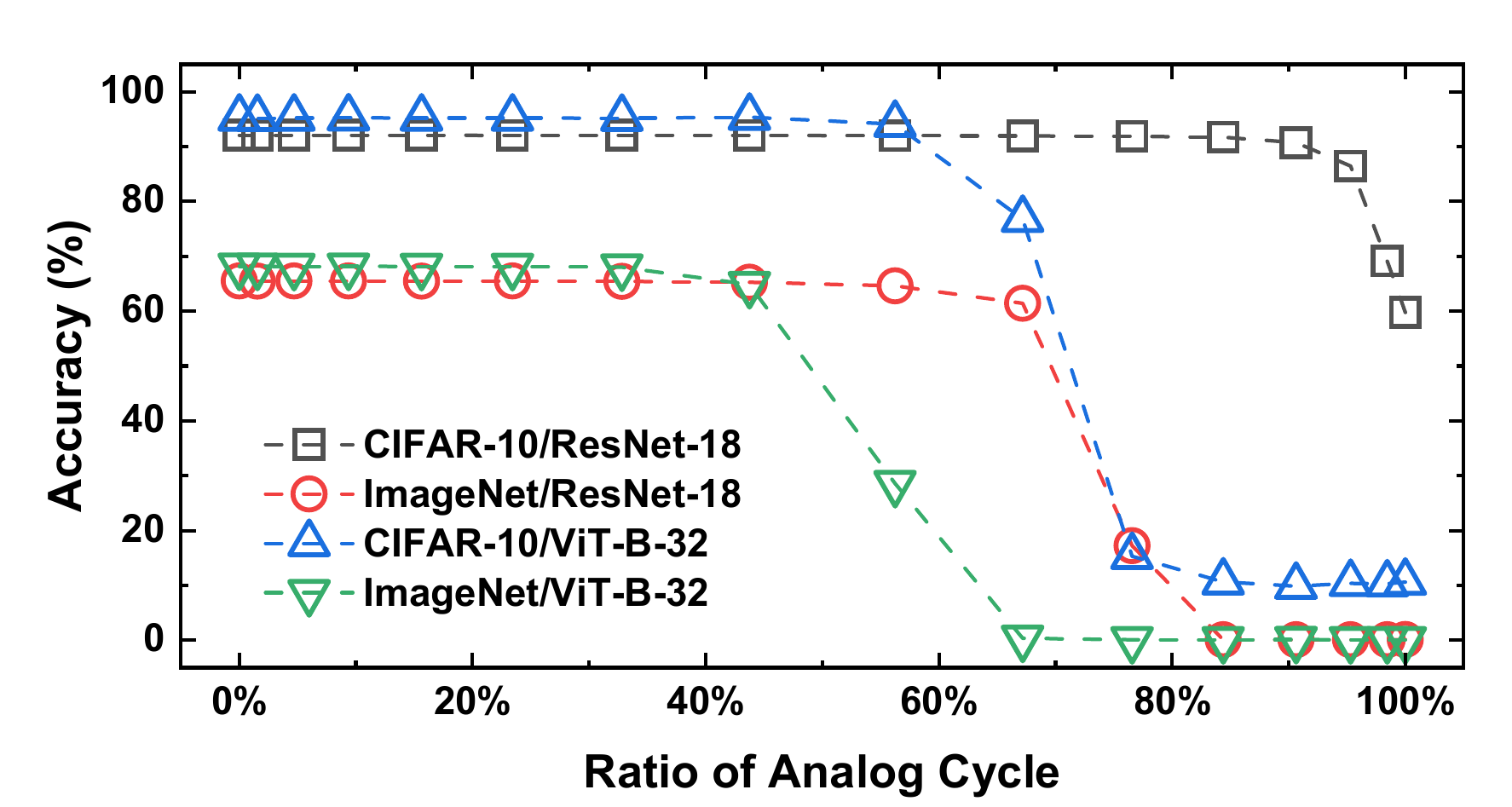}
        \label{Fig_HCiM_b}
    }%
    
    \caption{Digital-Analog HCiM for DNN inference accuracy improvement. (a) 8b/8b HCiM workload management and (b) Inference accuracy versus analog cycle ratio for different models and datasets using a noisy ACiM macro (noise intensity = 0.8 $\text{LSB}_{\text{rms}}$). A majority of cycles can be processed by ACiM, with MSB cycles allocated to the digital domain to achieve a balance between accuracy and efficiency.}
    \label{Fig_HCiM}
\end{figure}

Hybrid Analog-Digital computing is proposed to strike a balance between computing accuracy and efficiency, enabling both high accuracy and energy efficiency~\cite{Hybrid_Analog-Digital}. In the CiM domain, Analog-Digital HCiM has shown significant promise for practical applications~\cite{FP_HCIM,OSA-HCIM,PACiM}. HCiM separates binary cycles between DCiM and ACiM: the majority of cycles remain in the analog domain to maximize energy efficiency, while MSB cycles, which carry greater weight due to bit shifting, are processed in digital domain to maintain error within a manageable range, as depicted in Fig.~\ref{Fig_HCiM}(a). We evaluated HCiM accuracy by configuring a 256-row-parallel ACiM macro with an 8-bit ADC under high noise intensity (0.8 $\text{LSB}_{\text{rms}}$). Fig.~\ref{Fig_HCiM}(b) shows accuracy in relation to the analog cycle ratio, achieved by adjusting the analog-digital boundary across models and datasets. With a boundary level set to 3 in ResNet-18/CIFAR-10 task, transferring only 6 cycles to DCiM restores inference accuracy close to baseline, while retaining 91\% of cycles in ACiM under noisy conditions. Although previous sections show that complex tasks are sensitive to analog noise, transferring over half of the binary cycles to DCiM allows accuracy to match baseline while improving the energy efficiency. Even for the most challenging case of ViT-B-32 on ImageNet, maintaining over 40\% of cycles in the analog domain is sufficient to stabilize accuracy even in a less optimized ACiM macro.

ACiM begins to outperform DCiM at a row-parallelism of 64, with energy efficiency improving to a $7\times$ gain as row-parallelism scales up to 1024~\cite{Analog_or_Digital}. At a row-parallelism of 256, ACiM achieves approximately $3\times$ the TOPS/W of DCiM. Building upon this benchmark, the HCiM approach illustrated in Fig.~\ref{Fig_HCiM}(b) results in 6\%, 20\%, 30\%, and 36\% decrease in TOPS/W for CIFAR-10/ResNet-18, ImageNet/ResNet-18, CIFAR-10/ViT-B-32, and ImageNet/ViT-B-32, respectively, compared to full ACiM execution. However, HCiM still provides 182\%, 140\%, 110\%, and 90\% TOPS/W gains over full DCiM execution, showcasing its strong potential for energy-efficient CiM deployment in real-world applications.

\begin{table*}[htbp]
\centering
\caption{Comparison of Different Simulation Frameworks for DNN Inference Assessment.}
\label{Table_Comparison_Frameworks}
\begin{tabular}{c|ccccc}
\toprule
 & \textbf{CIMUFAS~\cite{Fast_Accurate_Sim}} & \textbf{Saikia's Modeling~\cite{Modeling_SRAM_IMC}} & \textbf{MLP+NeuroSim~\cite{NeuroSim}} & \textbf{DNN+NeuroSim~\cite{DNN+NeuroSim}} & \textbf{ASiM} \\
\midrule
\textbf{Primary Memory}   & RRAM & SRAM & RRAM & SRAM & SRAM \\
\textbf{Primary Domain}   & Current & Charge & Current & Current & Charge \\
\textbf{ADC Modeling}   & Gaussian & Adaptive & Min-Max & Min-Max & Full Dynamic Range \\
\textbf{CNN Support}   & $\checkmark$ & $\checkmark$ & $\times$ & $\checkmark$ & $\checkmark$ \\
\textbf{Transformer Support}   & $\times$ & $\times$ & $\times$ & $\times$ & $\checkmark$ \\
\textbf{Bit-Parallel Support}   & $\times$ & $\times$ & $\times$ & $\times$ & $\checkmark$ \\
\textbf{Modular Design Support}   & $\times$ & $\times$ & $\times$ & $\checkmark$ & $\checkmark$ \\
\bottomrule
\end{tabular}
\end{table*}

\subsection{Majority Voting}

\begin{figure}[htbp]
    \centering
    \subfloat[]{%
        \includegraphics[width=0.98\linewidth]{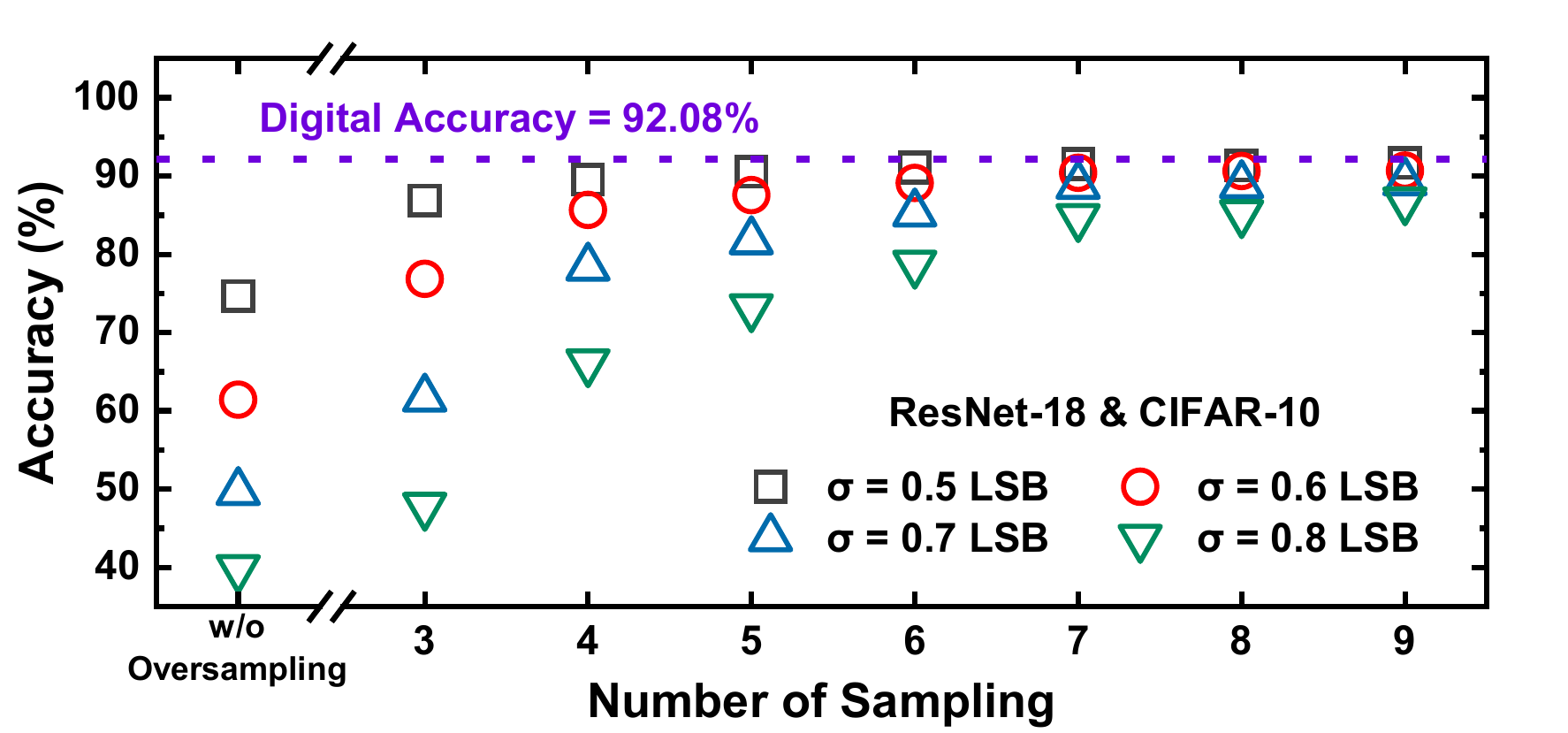}
        \label{Fig_Maj_Vot_a}
    }\\

    \subfloat[]{%
        \includegraphics[width=0.98\linewidth]{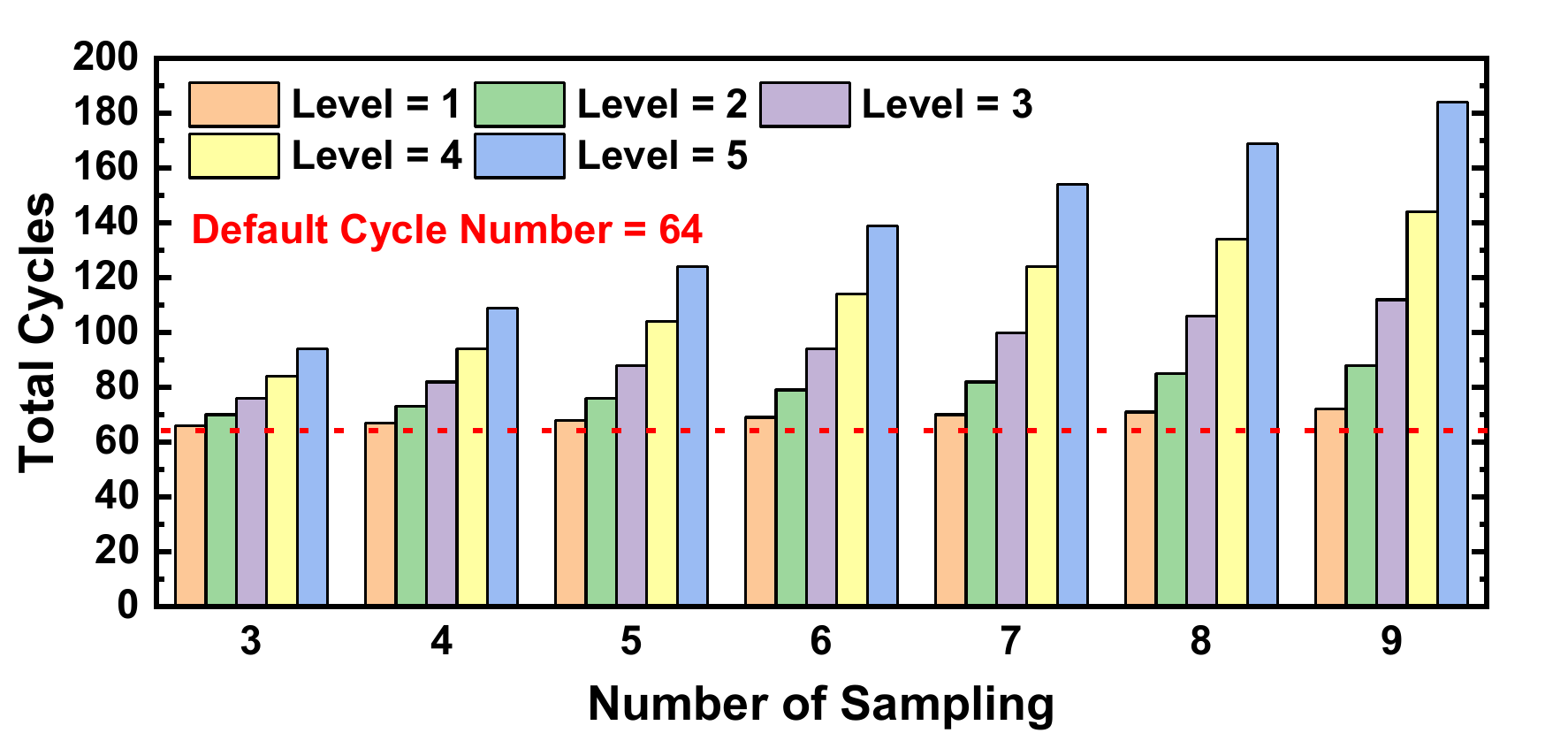}
        \label{Fig_Maj_Vot_b}
    }%
    
    \caption{Oversampling MSB cycles with majority voting to improve  inference accuracy. (a) Inference accuracy of ResNet-18/CIFAR-10 versus sampling count at varying analog noise intensities, with a boundary level set to 3. (b) Total cycle count across majority voting configurations, with the standard 8b/8b model requiring 64 bit-serial cycles.}
    \label{Fig_Maj_Vot}
\end{figure}

HCiM provides a way to fully eliminate inaccuracies in MSB cycles but requires additional digital logic in the ACiM arrays, adding complexity to the design. An alternative approach, majority voting, preserves the ACiM architecture but increases power consumption in exchange for enhanced inference accuracy. To achieve performance comparable to HCiM, we oversampled the MSB cycles in ACiM and applied majority voting to these MAC results. For the 6 MSB cycles (boundary level = 3) in the ResNet-18/CIFAR-10 task, we conducted oversampling at different noise intensities. As shown in Fig.~\ref{Fig_Maj_Vot}(a), the equivalent standard deviation of noise intensity decreases at a rate of $\sqrt{N}$~\cite{Oversampling_ADC}, and inference accuracy improves with increasing sampling times. Fig.~\ref{Fig_Maj_Vot}(b) shows the computational overhead associated with majority voting of an 8b/8b DNN model under different boundary level configurations and oversampling iterations. Even with a less optimized ACiM macro at a noise intensity of 0.7 $\text{LSB}_{\text{rms}}$, oversampling the top 6 cycles 7 times achieves 90\% inference accuracy, with a total cycle increase of less than 40\%. As digital accelerators generally consume significantly more power, ACiM retains its superiority over other alternative DNN accelerators. However, for complex tasks, designs must prioritize minimizing analog noise-induced errors. It is important to note that majority voting effectively reduces random noise, including kT/C noise and random comparator decision errors, but does not mitigate systematic errors. To compensate for systematic errors, approaches such as NAT or ADC code post-calibration are necessary to enhance inference performance.

\section{Discussion}
\label{Chapt_Discussion}

\subsection{Comparison with Other Simulation Frameworks}

Table~\ref{Table_Comparison_Frameworks} compares our ASiM framework with other inference simulation frameworks. One key distinction lies in the modeling of ADC behavior, which significantly impacts inference performance. Saikia’s modeling~\cite{Modeling_SRAM_IMC} employs an adaptive dynamic range, which is challenging to implement in real-world applications, whereas CIMUFAS~\cite{Fast_Accurate_Sim} bypasses ADC rounding by introducing Gaussian noise instead. NeuroSim~\cite{NeuroSim,DNN+NeuroSim} assumes that the ADC dynamic range is defined by the Min-Max range of each MAC output, leading to overly optimistic inference results. Unlike these frameworks, ASiM simulates ADC behavior across its full dynamic range, accurately capturing both rounding effects and noise-induced errors based on the real MAC output distribution. Additionally, ASiM supports the latest Transformer-based models and incorporates advanced features such as bit-parallel ACiM evaluation. Lastly, ASiM’s modular design allows for seamless integration into various codebases, making it a user-friendly and adaptable tool for ACiM circuit evaluation.

\subsection{Model Extension}

As ACiM technologies rapidly evolve, the incorporation of emerging memories and novel circuit designs continues to enhance system efficiency. Although ASiM is initially designed for charge-domain SRAM ACiM, its modular and adaptable structure allows users to easily modify it for other architectures. Taking RRAM as an example, where crossbar cells store multi-bit weights, the loop over binary weight bits in Algorithm~\ref{Algorithm_ASiM} is unnecessary and can be omitted. This adjustment enables each ADC to translate the quantized current value for a RRAM column. Meanwhile, the capacitor nonlinearity model used for SRAM should be replaced with a noise model that more accurately captures conductance variation in RRAM. With such minimal changes, users can leverage ASiM’s core logic to create more faithful simulations for their specific circuit implementations. We encourage the research community to extend ASiM to better support diverse and emerging ACiM technologies.

\subsection{Comparison Across State-of-the-Art CiM Designs}

\begin{figure}[htbp]
    \centering
    \includegraphics[width=\columnwidth]{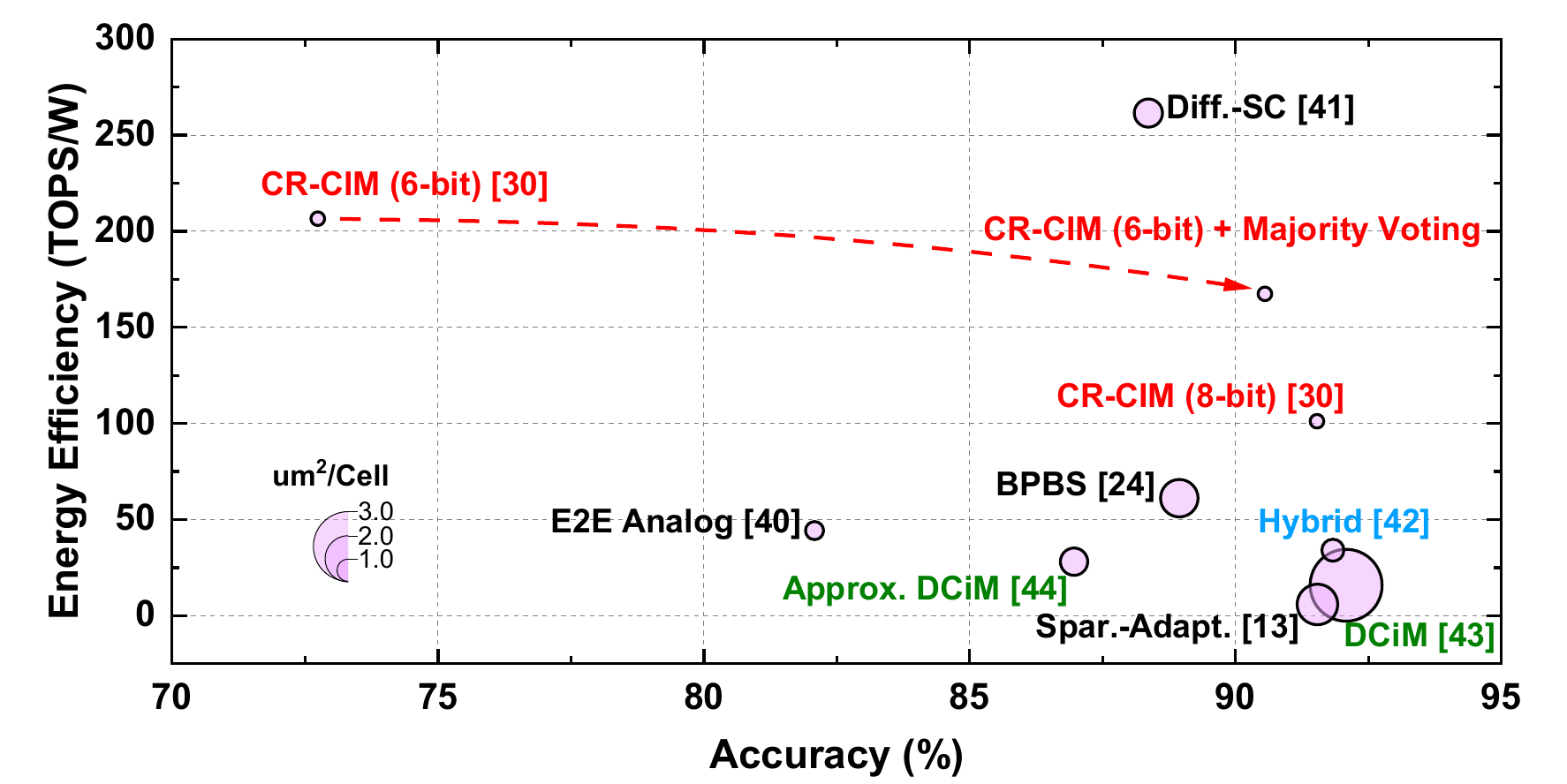}
    \caption{Comparison of state-of-the-art CiM designs analyzed using ASiM. Design parameters are listed in Table~\ref{Table_Design_Comp}. Bubble size indicates the estimated chip area, and both energy efficiency and area are normalized to a 0.9 V supply voltage at the 22 nm technology node using spatial scaling factors.}
    \label{Fig_Design_Comp}
\end{figure}

To evaluate the practical inference performance of SRAM-based ACiM, we compare representative designs that can be accurately modeled within the ASiM framework. All simulations are conducted using a unified 8b/8b ResNet-18 model on the CIFAR-10 benchmark. The resulting inference accuracy, energy efficiency, and area efficiency are normalized to a 22 nm technology node and 0.9 V supply voltage and summarized in Fig.~\ref{Fig_Design_Comp}. The size of each bubble represents the estimated area efficiency, calculated as the average area per basic CiM cell. Relevant design parameters are extracted from their respective publications and summarized in Table~\ref{Table_Design_Comp}. For designs where $\text{LSB}_{\text{rms}}$ was not reported, a nominal value of 0.5 LSB was assumed. The designs under evaluation include a range of representative schemes: the standard Bit-Parallel/Bit-Serial (\textbf{BPBS~\cite{Programmable_BSBP}}) architecture for multi-bit DNN processing; the Sparsity-Adaptive (\textbf{Spar.-Adapt.~\cite{Bit-Flexible_CIM}}) scheme, which skips redundant ADC conversions by exploiting activation sparsity; the End-to-End Analog (\textbf{E2E Analog~\cite{Adaptive-Ranging}}) pipeline, which maintains analog signal integrity throughout computation; the Differential Switched-Capacitor (\textbf{Diff.-SC~\cite{Differential_SC}}) design, which utilizes differential compute lines to enhance dynamic range and ADC precision; and the Capacitor-Reconfigured CiM (\textbf{CR-CIM~\cite{CR_ACIM_ISSCC}}) scheme, which reuses cell capacitors to improve ADC resolution. In addition, CR-CIM is also configured with a reduced 6-bit ADC to demonstrate accuracy degradation, followed by recovery via majority voting. For completeness, the HCiM (\textbf{Hybrid~\cite{Bit-Rotated}}) and two DCiM designs~\cite{DCiM,DIMC} are also included as references.

\begin{table}[htbp]
\centering
\caption{Parameters of state-of-the-art ACiM designs presented in Fig.~\ref{Fig_Design_Comp}, with specifications extracted from their publications.}
\label{Table_Design_Comp}
\begin{tabular}{c|cccc}
\toprule
\textbf{Design} & \textbf{Parallelism} & \textbf{Act. Enc.} & \textbf{ADC} & $\textbf{LSB}_{\textbf{rms}}$ \\
\midrule
\textbf{BPBS~\cite{Programmable_BSBP}} & 2,305 & 1b & 8b & 0.37 \\
\textbf{Spar.-Adapt.~\cite{Bit-Flexible_CIM}}   & 256 & 1b & 8b & N/A \\
\textbf{E2E Analog~\cite{Adaptive-Ranging}}   & 64 & 4b & 5b & 0.27 \\
\textbf{Diff.-SC~\cite{Differential_SC}}   & 1,152 & 4b & 10b & 0.40 \\
\textbf{CR-CIM~\cite{CR_ACIM_ISSCC}}   & 512 & 1b & 10b & 0.58 \\
\textbf{Hybrid~\cite{Bit-Rotated}}   & 256 & 2b & 8b & N/A \\
\bottomrule
\end{tabular}
\end{table}

Based on Fig.~\ref{Fig_Design_Comp}, the \textbf{Spar.-Adapt.} and 8-bit \textbf{CR-CIM} designs yield the highest inference accuracy, primarily because they operate above the boundary ADC precision. In contrast, more aggressive designs typically demonstrate superior energy efficiency but at the cost of reduced inference accuracy. Designs with lower reported $\text{LSB}_{\text{rms}}$ values tend to better support such aggressive configurations, underscoring the importance of meticulous ACiM design. The HCiM architecture proposed in \cite{Bit-Rotated} achieves approximately $2\times$ energy efficiency improvement and significantly reduced area cost compared to standard DCiM, while incurring negligible accuracy loss, highlighting its promise for future exploration. Furthermore, by applying 3 times oversampling to majority voting with a boundary level of 3, the 6-bit \textbf{CR-CIM} design recovers accuracy from 72.74\% to 90.55\%, while achieving a 65\% improvement in energy efficiency compared to its 8-bit configuration. This indicates its strong potential for low-power applications where moderate accuracy degradation is acceptable.

\section{Conclusions}
\label{Chapt_Conclusion}

We presented ASiM in this paper, an open-source simulation framework specifically developed to assess and optimize inference accuracy in SRAM-based ACiM circuits. ASiM allows researchers to easily integrate their ACiM circuits into the PyTorch ecosystem, enabling the evaluation of various design factors under realistic conditions that include quantization noise, analog imperfections, and ADC constraints. Our extensive analysis using ASiM revealed critical insights for ACiM design. Although the bit-parallel approach can substantially improve energy efficiency while tolerating quantization noise, analog noise poses a serious threat to inference accuracy, creating stringent design requirements. We also showed that higher ADC precision is crucial for more complex models and tasks, such as ImageNet and ViT, while CNNs on simpler datasets can reduce ADC precision without significant accuracy loss. Furthermore, we found that largest computation errors primarily occur in the MSB cycles, where even a minor ADC readout error as small as 1 LSB can result in severe inference degradation. To tackle these challenges, we recommended two strategies: HCiM and majority voting, both of which aim to ensure accurate MSB cycle computation. Incorporating these methods shows promise in enhancing inference accuracy without diminishing inherent advantage on energy efficiency, presenting a promising strategy for deploying ACiM reliably in real-world applications.

% \section*{Acknowledgments}

\bibliographystyle{IEEEtran}
\bibliography{reference}

\begin{thebibliography}{10}
\providecommand{\url}[1]{#1}
\csname url@samestyle\endcsname
\providecommand{\newblock}{\relax}
\providecommand{\bibinfo}[2]{#2}
\providecommand{\BIBentrySTDinterwordspacing}{\spaceskip=0pt\relax}
\providecommand{\BIBentryALTinterwordstretchfactor}{4}
\providecommand{\BIBentryALTinterwordspacing}{\spaceskip=\fontdimen2\font plus
\BIBentryALTinterwordstretchfactor\fontdimen3\font minus \fontdimen4\font\relax}
\providecommand{\BIBforeignlanguage}[2]{{%
\expandafter\ifx\csname l@#1\endcsname\relax
\typeout{** WARNING: IEEEtran.bst: No hyphenation pattern has been}%
\typeout{** loaded for the language `#1'. Using the pattern for}%
\typeout{** the default language instead.}%
\else
\language=\csname l@#1\endcsname
\fi
#2}}
\providecommand{\BIBdecl}{\relax}
\BIBdecl

\bibitem{LLM_Scaling}
J.~Kaplan, S.~McCandlish, T.~Henighan, T.~B. Brown, B.~Chess, R.~Child, S.~Gray, A.~Radford, J.~Wu, and D.~Amodei, ``Scaling laws for neural language models,'' \emph{ArXiv}, vol. abs/2001.08361, 2020.

\bibitem{Energy_Problem}
M.~Horowitz, ``1.1 computing's energy problem (and what we can do about it),'' in \emph{2014 IEEE International Solid-State Circuits Conference Digest of Technical Papers (ISSCC)}, 2014, pp. 10--14.

\bibitem{IMC_Advances_Prospects}
N.~Verma, H.~Jia, H.~Valavi, Y.~Tang, M.~Ozatay, L.-Y. Chen, B.~Zhang, and P.~Deaville, ``In-memory computing: Advances and prospects,'' \emph{IEEE Solid-State Circuits Magazine}, vol.~11, no.~3, pp. 43--55, 2019.

\bibitem{Analog_or_Digital}
J.~Sun, P.~Houshmand, and M.~Verhelst, ``Analog or digital in-memory computing? benchmarking through quantitative modeling,'' in \emph{2023 IEEE/ACM International Conference on Computer Aided Design (ICCAD)}, 2023, pp. 1--9.

\bibitem{64-Tile}
H.~Valavi, P.~J. Ramadge, E.~Nestler, and N.~Verma, ``A 64-tile 2.4-mb in-memory-computing cnn accelerator employing charge-domain compute,'' \emph{IEEE Journal of Solid-State Circuits}, vol.~54, no.~6, pp. 1789--1799, 2019.

\bibitem{MACC-SRAM}
B.~Zhang, J.~Saikia, J.~Meng, D.~Wang, S.~Kwon, S.~Myung, H.~Kim, S.~J. Kim, J.-S. Seo, and M.~Seok, ``Macc-sram: A multistep accumulation capacitor-coupling in-memory computing sram macro for deep convolutional neural networks,'' \emph{IEEE Journal of Solid-State Circuits}, vol.~59, no.~6, pp. 1938--1949, 2024.

\bibitem{In-SRAM-ADC_ACIM}
K.~Lee, J.~Kim, and J.~Park, ``A 28-nm 50.1-tops/w p-8t sram compute-in-memory macro design with bl charge-sharing-based in-sram dac/adc operations,'' \emph{IEEE Journal of Solid-State Circuits}, vol.~59, no.~6, pp. 1926--1937, 2024.

\bibitem{PIMCA}
B.~Zhang, S.~Yin, M.~Kim, J.~Saikia, S.~Kwon, S.~Myung, H.~Kim, S.~J. Kim, J.-S. Seo, and M.~Seok, ``Pimca: A programmable in-memory computing accelerator for energy-efficient dnn inference,'' \emph{IEEE Journal of Solid-State Circuits}, vol.~58, no.~5, pp. 1436--1449, 2023.

\bibitem{One-Shot_AD_MAC}
X.~Yang and N.~Sun, ``A 4-bit mixed-signal mac macro with one-shot adc conversion,'' \emph{IEEE Journal of Solid-State Circuits}, vol.~58, no.~9, pp. 2648--2658, 2023.

\bibitem{PICO-RAM}
Z.~Chen, Z.~Wen, W.~Wan, A.~R. Pakala, Y.~Zou, W.-C. Wei, Z.~Li, Y.~Chen, and K.~Yang, ``Pico-ram: A pvt-insensitive analog compute-in-memory sram macro with in situ multi-bit charge computing and 6t thin-cell-compatible layout,'' \emph{IEEE Journal of Solid-State Circuits}, pp. 1--13, 2024.

\bibitem{Programmable_Accelerator_ISSCC}
H.~Jia, M.~Ozatay, Y.~Tang, H.~Valavi, R.~Pathak, J.~Lee, and N.~Verma, ``15.1 a programmable neural-network inference accelerator based on scalable in-memory computing,'' in \emph{2021 IEEE International Solid-State Circuits Conference (ISSCC)}, vol.~64, 2021, pp. 236--238.

\bibitem{C3SRAM}
Z.~Jiang, S.~Yin, J.-S. Seo, and M.~Seok, ``C3sram: An in-memory-computing sram macro based on robust capacitive coupling computing mechanism,'' \emph{IEEE Journal of Solid-State Circuits}, vol.~55, no.~7, pp. 1888--1897, 2020.

\bibitem{Bit-Flexible_CIM}
C.-Y. Yao, T.-Y. Wu, H.-C. Liang, Y.-K. Chen, and T.-T. Liu, ``A fully bit-flexible computation in memory macro using multi-functional computing bit cell and embedded input sparsity sensing,'' \emph{IEEE Journal of Solid-State Circuits}, vol.~58, no.~5, pp. 1487--1495, 2023.

\bibitem{NeuroSim}
P.-Y. Chen, X.~Peng, and S.~Yu, ``Neurosim: A circuit-level macro model for benchmarking neuro-inspired architectures in online learning,'' \emph{IEEE Transactions on Computer-Aided Design of Integrated Circuits and Systems}, vol.~37, no.~12, pp. 3067--3080, 2018.

\bibitem{SySCIM}
S.~H.~H. Shadmehri, A.~BanaGozar, M.~Kamal, S.~Stuijk, A.~Afzali-Kusha, M.~Pedram, and H.~Corporaal, ``Syscim: Systemc-ams simulation of memristive computation in-memory,'' in \emph{2022 Design, Automation \& Test in Europe Conference \& Exhibition (DATE)}, 2022, pp. 1467--1472.

\bibitem{PIMSIM-NN}
X.~Wang, X.~Sun, Y.~Han, and X.~Chen, ``Pimsim-nn: An isa-based simulation framework for processing-in-memory accelerators,'' in \emph{2024 Design, Automation \& Test in Europe Conference \& Exhibition (DATE)}, 2024, pp. 1--2.

\bibitem{aihwkit}
M.~J. Rasch, D.~Moreda, T.~Gokmen, M.~Le~Gallo, F.~Carta, C.~Goldberg, K.~El~Maghraoui, A.~Sebastian, and V.~Narayanan, ``A flexible and fast pytorch toolkit for simulating training and inference on analog crossbar arrays,'' in \emph{2021 IEEE 3rd International Conference on Artificial Intelligence Circuits and Systems (AICAS)}, 2021, pp. 1--4.

\bibitem{MemTorch}
C.~Lammie, W.~Xiang, B.~Linares-Barranco, and M.~{Rahimi Azghadi}, ``Memtorch: An open-source simulation framework for memristive deep learning systems,'' \emph{Neurocomputing}, vol. 485, pp. 124--133, 2022.

\bibitem{eF2lowSim}
J.~Wang, S.~Kim, J.~Heo, and C.~S. Park, ``ef2lowsim: System-level simulator of eflash-based compute-in-memory accelerators for convolutional neural networks,'' in \emph{2023 Design, Automation \& Test in Europe Conference \& Exhibition (DATE)}, 2023, pp. 1--6.

\bibitem{Fast_Accurate_Sim}
Z.~Wang, J.~Yue, C.~He, Z.~Dai, F.~Xiang, Z.~Cong, Y.~He, X.~Feng, and Y.~Liu, ``A user-friendly fast and accurate simulation framework for non-ideal factors in computing-in-memory architecture,'' in \emph{2023 IEEE International Symposium on Circuits and Systems (ISCAS)}, 2023, pp. 1--5.

\bibitem{X-PIM}
I.~Jeong and J.-E. Park, ``X-pim: Fast modeling and validation framework for mixed-signal processing-in-memory using compressed equivalent model in system verilog,'' in \emph{2024 Design, Automation \& Test in Europe Conference \& Exhibition (DATE)}, 2024, pp. 1--6.

\bibitem{Accuracy_Efficiency_Trade_Off}
M.~Kang, Y.~Kim, A.~D. Patil, and N.~R. Shanbhag, ``Deep in-memory architectures for machine learning–accuracy versus efficiency trade-offs,'' \emph{IEEE Transactions on Circuits and Systems I: Regular Papers}, vol.~67, no.~5, pp. 1627--1639, 2020.

\bibitem{CSNR_Metrics}
S.~K. Gonugondla, C.~Sakr, H.~Dbouk, and N.~R. Shanbhag, ``Fundamental limits on the precision of in-memory architectures,'' in \emph{2020 IEEE/ACM International Conference On Computer Aided Design (ICCAD)}, 2020, pp. 1--9.

\bibitem{Programmable_BSBP}
H.~Jia, H.~Valavi, Y.~Tang, J.~Zhang, and N.~Verma, ``A programmable heterogeneous microprocessor based on bit-scalable in-memory computing,'' \emph{IEEE Journal of Solid-State Circuits}, vol.~55, no.~9, pp. 2609--2621, 2020.

\bibitem{CR_ACIM_JSSC}
K.~Yoshioka, ``A 818–4094 tops/w capacitor-reconfigured analog cim for unified acceleration of cnns and transformers,'' \emph{IEEE Journal of Solid-State Circuits}, pp. 1--12, 2024.

\bibitem{Noise_Modeling_SAR}
W.~P. Zhang and X.~Tong, ``Noise modeling and analysis of sar adcs,'' \emph{IEEE Transactions on Very Large Scale Integration (VLSI) Systems}, vol.~23, no.~12, pp. 2922--2930, 2015.

\bibitem{Random_Decision_Error}
J.~Kim, B.~S. Leibowitz, J.~Ren, and C.~J. Madden, ``Simulation and analysis of random decision errors in clocked comparators,'' \emph{IEEE Transactions on Circuits and Systems I: Regular Papers}, vol.~56, no.~8, pp. 1844--1857, 2009.

\bibitem{Cap_Non_linear}
V.~Tripathi and B.~Murmann, ``Mismatch characterization of small metal fringe capacitors,'' \emph{IEEE Transactions on Circuits and Systems I: Regular Papers}, vol.~61, no.~8, pp. 2236--2242, 2014.

\bibitem{Modeling_SRAM_IMC}
J.~Saikia, S.~Yin, S.~K. Cherupally, B.~Zhang, J.~Meng, M.~Seok, and J.-S. Seo, ``Modeling and optimization of sram-based in-memory computing hardware design,'' in \emph{2021 Design, Automation \& Test in Europe Conference \& Exhibition (DATE)}, 2021, pp. 942--947.

\bibitem{CR_ACIM_ISSCC}
K.~Yoshioka, ``34.5 a 818-4094tops/w capacitor-reconfigured cim macro for unified acceleration of cnns and transformers,'' in \emph{2024 IEEE International Solid-State Circuits Conference (ISSCC)}, vol.~67, 2024, pp. 574--576.

\bibitem{DNN+NeuroSim}
X.~Peng, S.~Huang, Y.~Luo, X.~Sun, and S.~Yu, ``Dnn+neurosim: An end-to-end benchmarking framework for compute-in-memory accelerators with versatile device technologies,'' in \emph{2019 IEEE International Electron Devices Meeting (IEDM)}, 2019, pp. 32.5.1--32.5.4.

\bibitem{White_Paper_Quantization}
M.~Nagel, M.~Fournarakis, R.~A. Amjad, Y.~Bondarenko, M.~van Baalen, and T.~Blankevoort, ``A white paper on neural network quantization,'' \emph{ArXiv}, vol. abs/2106.08295, 2021.

\bibitem{Low-Cost_7T}
K.~Lee, J.~Kim, and J.~Park, ``Low-cost 7t-sram compute-in-memory design based on bit-line charge-sharing based analog-to-digital conversion,'' in \emph{2022 IEEE/ACM International Conference On Computer Aided Design (ICCAD)}, 2022, pp. 1--8.

\bibitem{Challenges_Transformer_Quantization}
Y.~Bondarenko, M.~Nagel, and T.~Blankevoort, ``Understanding and overcoming the challenges of efficient transformer quantization,'' \emph{ArXiv}, vol. abs/2109.12948, 2021.

\bibitem{PACiM}
W.~Zhang, S.~Ando, Y.-C. Chen, S.~Miyagi, S.~Takamaeda-Yamazaki, and K.~Yoshioka, ``Pacim: A sparsity-centric hybrid compute-in-memory architecture via probabilistic approximation,'' in \emph{Proceedings of the 43rd IEEE/ACM International Conference on Computer-Aided Design}, 2024.

\bibitem{Hybrid_Analog-Digital}
M.~R. Haq~Rashed, S.~K. Jha, and R.~Ewetz, ``Hybrid analog-digital in-memory computing,'' in \emph{2021 IEEE/ACM International Conference On Computer Aided Design (ICCAD)}, 2021, pp. 1--9.

\bibitem{FP_HCIM}
P.-C. Wu, J.-W. Su, L.-Y. Hong, J.-S. Ren, C.-H. Chien, H.-Y. Chen, C.-E. Ke, H.-M. Hsiao, S.-H. Li, S.-S. Sheu, W.-C. Lo, S.-C. Chang, C.-C. Lo, R.-S. Liu, C.-C. Hsieh, K.-T. Tang, and M.-F. Chang, ``A 22nm 832kb hybrid-domain floating-point sram in-memory-compute macro with 16.2-70.2tflops/w for high-accuracy ai-edge devices,'' in \emph{2023 IEEE International Solid-State Circuits Conference (ISSCC)}, 2023, pp. 126--128.

\bibitem{OSA-HCIM}
Y.-C. Chen, S.~Ando, D.~Fujiki, S.~Takamaeda-Yamazaki, and K.~Yoshioka, ``Osa-hcim: On-the-fly saliency-aware hybrid sram cim with dynamic precision configuration,'' in \emph{2024 29th Asia and South Pacific Design Automation Conference (ASP-DAC)}, 2024, pp. 539--544.

\bibitem{Oversampling_ADC}
A.~Verreault, P.-V. Cicek, and A.~Robichaud, ``Oversampling adc: A review of recent design trends,'' \emph{IEEE Access}, vol.~12, pp. 121\,753--121\,779, 2024.

\bibitem{Adaptive-Ranging}
K.~Shiba, Z.~Zhan, K.~Nii, Y.~Wang, T.-Y.~J. Chang, A.~Kosuge, M.~Hamada, and T.~Kuroda, ``A 28-nm 0.8m-weights/mm2 9.1-tops/mm2 sram-based all-analog compute-in-memory using fine-grained structured pruning with adaptive-ranging adc,'' in \emph{2024 IEEE European Solid-State Electronics Research Conference (ESSERC)}, 2024, pp. 365--368.

\bibitem{Differential_SC}
J.~Lee, B.~Zhang, and N.~Verma, ``A switched-capacitor sram in-memory computing macro with high-precision, high-efficiency differential architecture,'' in \emph{2024 IEEE European Solid-State Electronics Research Conference (ESSERC)}, 2024, pp. 357--360.

\bibitem{Bit-Rotated}
X.~Chen, S.~Li, Z.~Zhang, W.~Zheng, X.~Tan, Y.~Tang, Y.~Shi, L.~Ren, Y.~Mai, F.~Liu, J.~Chen, Z.~Zhang, A.~Guo, T.~Xiong, B.~Wang, X.~Liu, W.~Shan, B.~Liu, H.~Cai, J.~Yang, and X.~Si, ``14.6 a 28nm 64kb bit-rotated hybrid-cim macro with an embedded sign-bit-processing array and a multi-bit-fusion dual-granularity cooperative quantizer,'' in \emph{2025 IEEE International Solid-State Circuits Conference (ISSCC)}, vol.~68, 2025, pp. 260--262.

\bibitem{DCiM}
Y.-D. Chih, P.-H. Lee, H.~Fujiwara, Y.-C. Shih, C.-F. Lee, R.~Naous, Y.-L. Chen, C.-P. Lo, C.-H. Lu, H.~Mori, W.-C. Zhao, D.~Sun, M.~E. Sinangil, Y.-H. Chen, T.-L. Chou, K.~Akarvardar, H.-J. Liao, Y.~Wang, M.-F. Chang, and T.-Y.~J. Chang, ``16.4 an 89tops/w and 16.3tops/mm2 all-digital sram-based full-precision compute-in memory macro in 22nm for machine-learning edge applications,'' in \emph{2021 IEEE International Solid-State Circuits Conference (ISSCC)}, vol.~64, 2021, pp. 252--254.

\bibitem{DIMC}
D.~Wang, C.-T. Lin, G.~K. Chen, P.~Knag, R.~K. Krishnamurthy, and M.~Seok, ``Dimc: 2219tops/w 2569f2/b digital in-memory computing macro in 28nm based on approximate arithmetic hardware,'' in \emph{2022 IEEE International Solid-State Circuits Conference (ISSCC)}, vol.~65, 2022, pp. 266--268.

\end{thebibliography}

% \newpage


% \end{IEEEbiography}



\vfill

\clearpage

\appendix

\begin{figure*}[htbp]
\centering
\includegraphics[width=\textwidth]{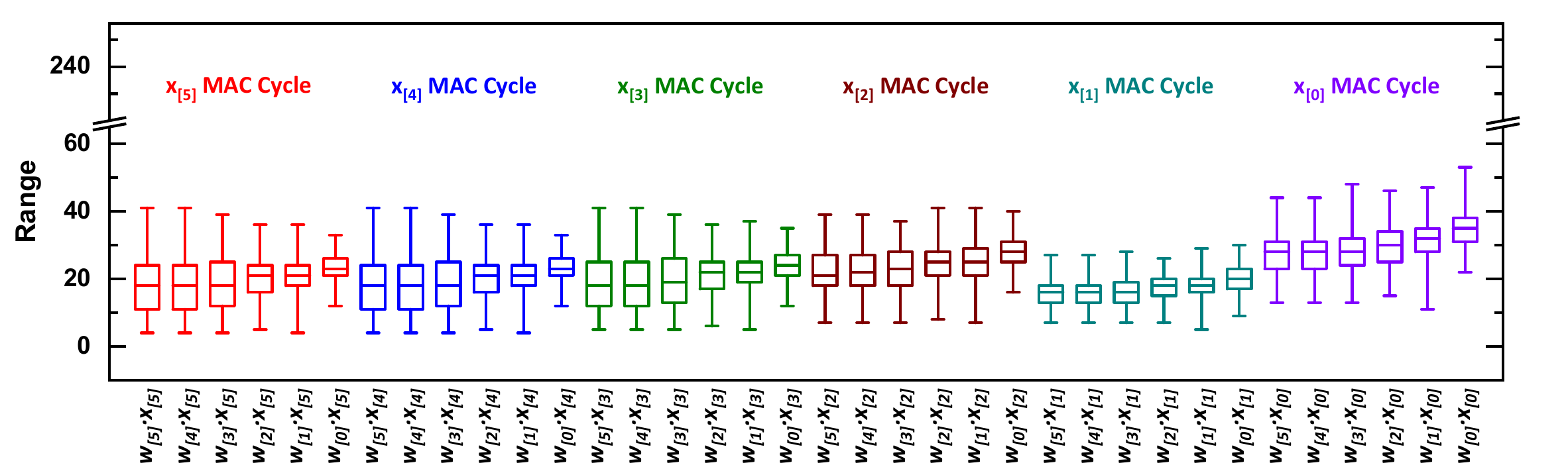}
\caption{Distribution of MAC outputs in the Query Projection of a 6b/6b ViT-B-32 for each bit-serial cycle, using an ACiM macro with row parallelism of 256 and 8-bit ADC.}
\label{Fig_Appendix_ViT_MAC}
\end{figure*}

\begin{figure}[htbp]
    \centering
    \subfloat[]{%
        \includegraphics[width=0.98\linewidth]{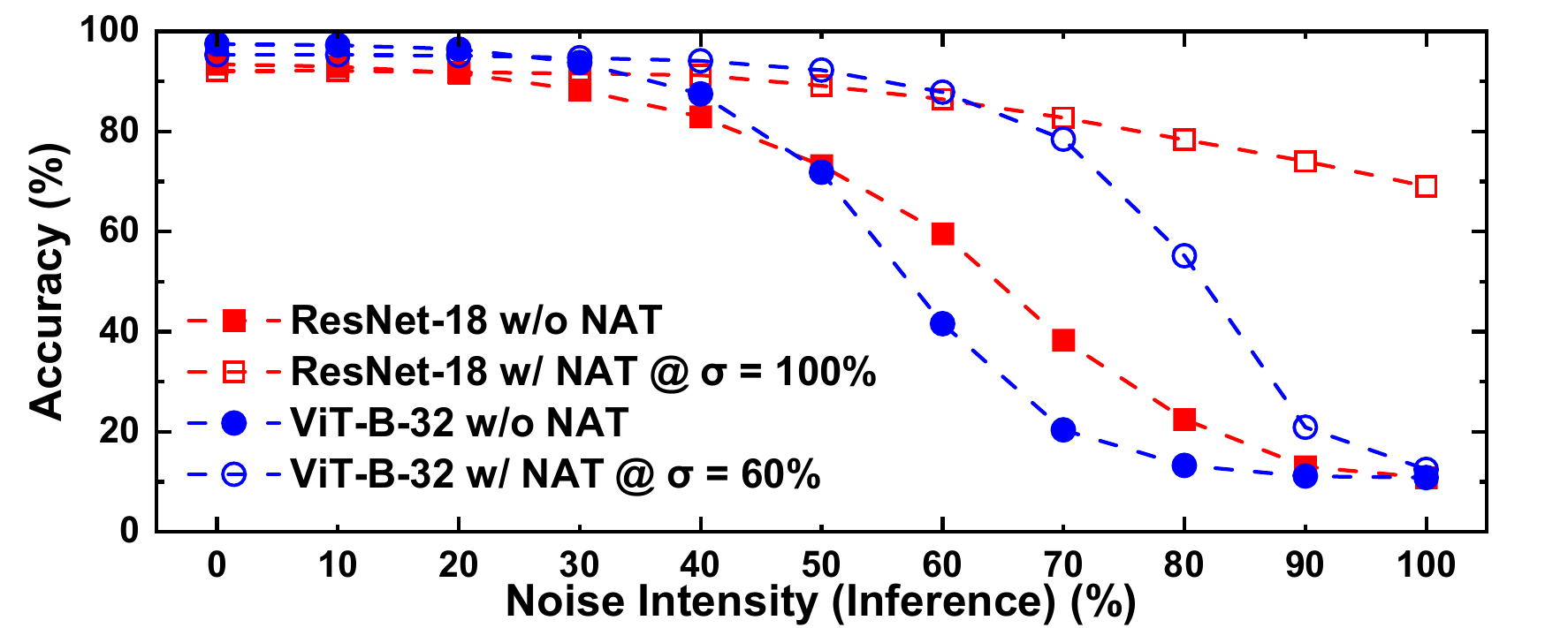}
        \label{}
    }\\
    \subfloat[]{%
        \includegraphics[width=0.98\linewidth]{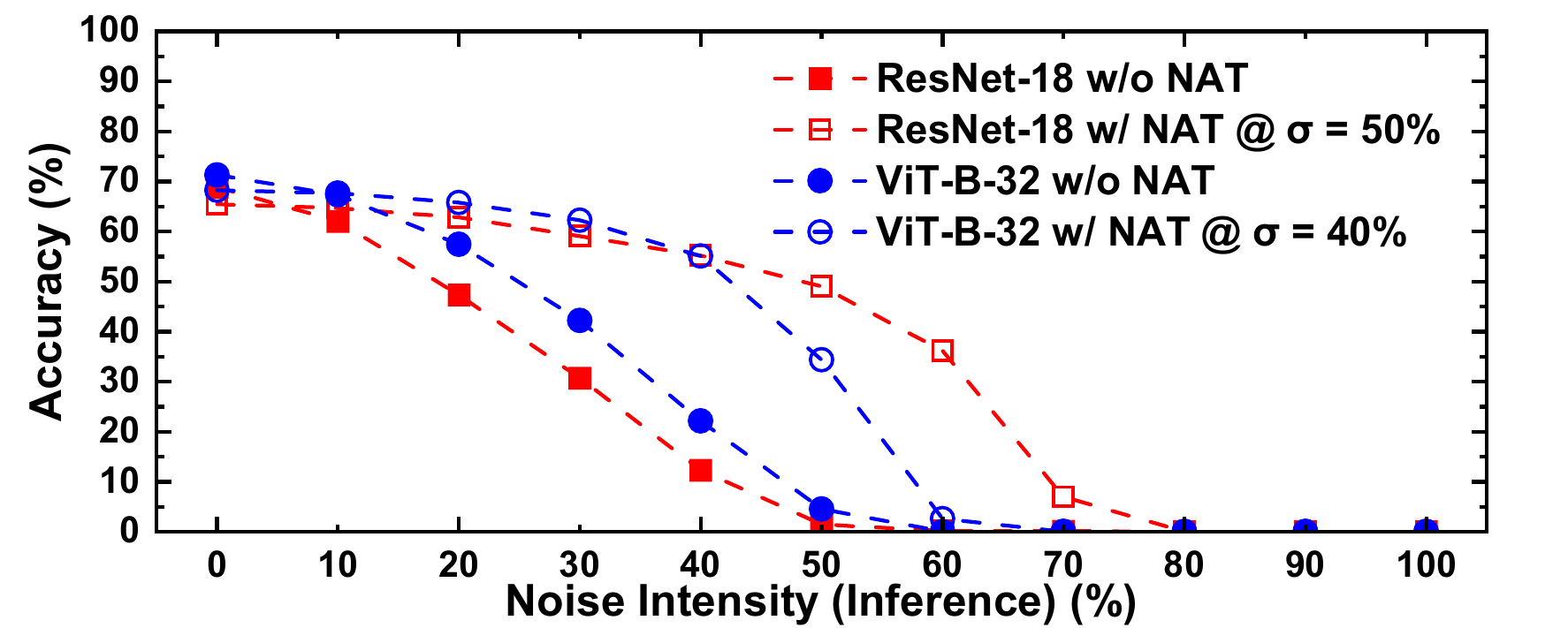}
        \label{}
    }%
    \caption{Inference accuracy of 8b/8b ResNet-18 and ViT-B/32 models, with and without NAT, evaluated under varying noise intensities. (a) CIFAR-10. (b) ImageNet.}
    \label{Fig_Appendix_NAT_Inference}
\end{figure}

\begin{figure}[htbp]
    \centering
    \includegraphics[width=\columnwidth]{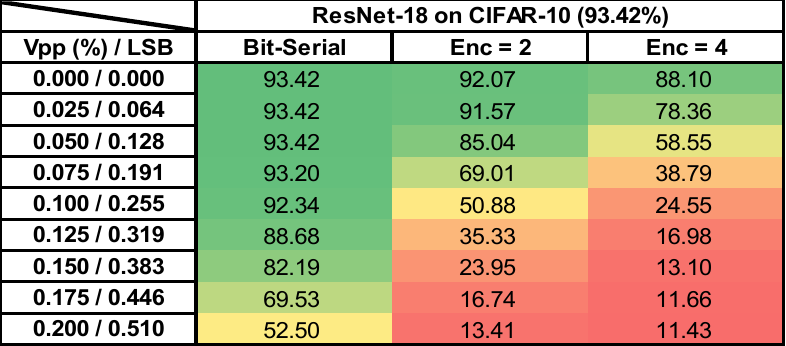}
    \caption{Shmoo plot of inference accuracy across different random noise intensities for an ACiM macro with a row-parallelism of 256 using 8-bit ADC, evaluated by an 8b/8b ResNet-18 on CIFAR-10.}
    \label{Fig_Appendix_Baseline_Sim}
\end{figure}

\section*{A General Perspective on MAC Distribution}

Although we only illustrate the MAC output of convolutional layers in CNNs, we argue that the trend shown in Fig.~\ref{Fig_MAC_Dist} is generalizable across diverse DNN models and layers. In particular, the MAC output of each ACiM column is proportional to the bit-level sparsity of weights and activations. The expected MAC value at an ACiM column can be expressed as:

\begin{equation}
P(DP = 1) = P(w = 1 \cap x = 1) = P(w = 1) \cdot P(x = 1),
\label{Eq_DP_Probability}
\end{equation}

\begin{equation}
E(MAC) = n \cdot P(DP = 1).
\label{Eq_Expect_MAC}
\end{equation}

\noindent Here, the expected MAC output equals the product of the row parallelism $n$ and the probability that a local dot product evaluates to "1". This probability further decomposes into the product of the bit-level sparsity of weights and activations~\cite{PACiM}. The derivation indicates that MAC outputs are fundamentally governed by sparsity patterns. For example, even when both weight and activation bits have a 50\% probability of being "1", the MAC output only spans roughly 25\% of the full dynamic range. Thus, sparsity inherently constrains the MAC span, and this trend persists across different depths and architectures.

To further analyze distribution characteristics, the weights and input activations in DNNs generally follow two representative types: \textbf{(1)} a symmetric, Gaussian-like distribution, or \textbf{(2)} a one-sided long-tailed distribution. After quantization into binary representations, their bit-level sparsity patterns align with those shown in Fig.~\ref{Fig_Input_Dist}(b). For type \textbf{(1)}, the bit sparsity across indices fluctuates around 50\%, whereas for type \textbf{(2)}, it typically falls between 0\% and 40\%, depending on the bit index. Although distribution shapes may vary across models and layers, the resulting bit-level sparsity after quantization remains consistent. As illustrated in Fig.~\ref{Fig_Input_Dist}(b), two dominant sparsity curves correspond to these two distribution types, despite some shape differences observed in Fig.\ref{Fig_Input_Dist}(a). Overall, modern DNN weights and activations largely conform to one of these two categories. Consequently, according to Eq.~\ref{Eq_Expect_MAC}, the column-wise MAC voltages observed by ADCs are confined within a limited dynamic range, which warrants careful modeling for inference accuracy.

For reference, we analyze the MAC outputs of a 256-row-parallel ACiM macro in the Query Projection of the Attention Block in ViT-B-32, with both weights and activations quantized to 6 bits. As shown in Fig.~\ref{Fig_Appendix_ViT_MAC}, the observed MAC values remain consistently below 60, which is far smaller than the theoretical maximum of 256. Remarkably, this distribution is even narrower than that observed in CNN layers (Fig.~\ref{Fig_MAC_Dist}), thereby imposing stricter precision requirements on ACiM circuits.

\section*{Quantitative Evaluation of Noise-Aware Training}

In this section, we quantitatively evaluate the effectiveness of NAT. We analyze ResNet-18 and ViT-B-32 models trained on CIFAR-10 and ImageNet with noise intensities of 100\%, 60\%, 50\%, and 40\%. These models were subsequently tested under varying inference-time noise levels and compared with their non-NAT counterparts, as shown in Fig.~\ref{Fig_Appendix_NAT_Inference}. The results clearly indicate that NAT enhances model robustness, enabling improved tolerance to higher noise levels during inference.

To strengthen our argument, we simulate the 8b/8b ResNet-18 model on CIFAR-10 using the same experimental configuration as in Fig.~\ref{Fig_Accuracy_Noise}(a), except that the weights are obtained without NAT. The corresponding accuracy shmoo in Fig.~\ref{Fig_Appendix_Baseline_Sim} shows a sharper decline relative to the NAT-enhanced counterpart, thereby underscoring the necessity of NAT in improving the robustness of ACiM-based systems.

\section*{Examining Noise Modeling in ASiM}

\begin{figure*}[t]
  \centering
  \subfloat[]{\includegraphics[width=0.32\textwidth]{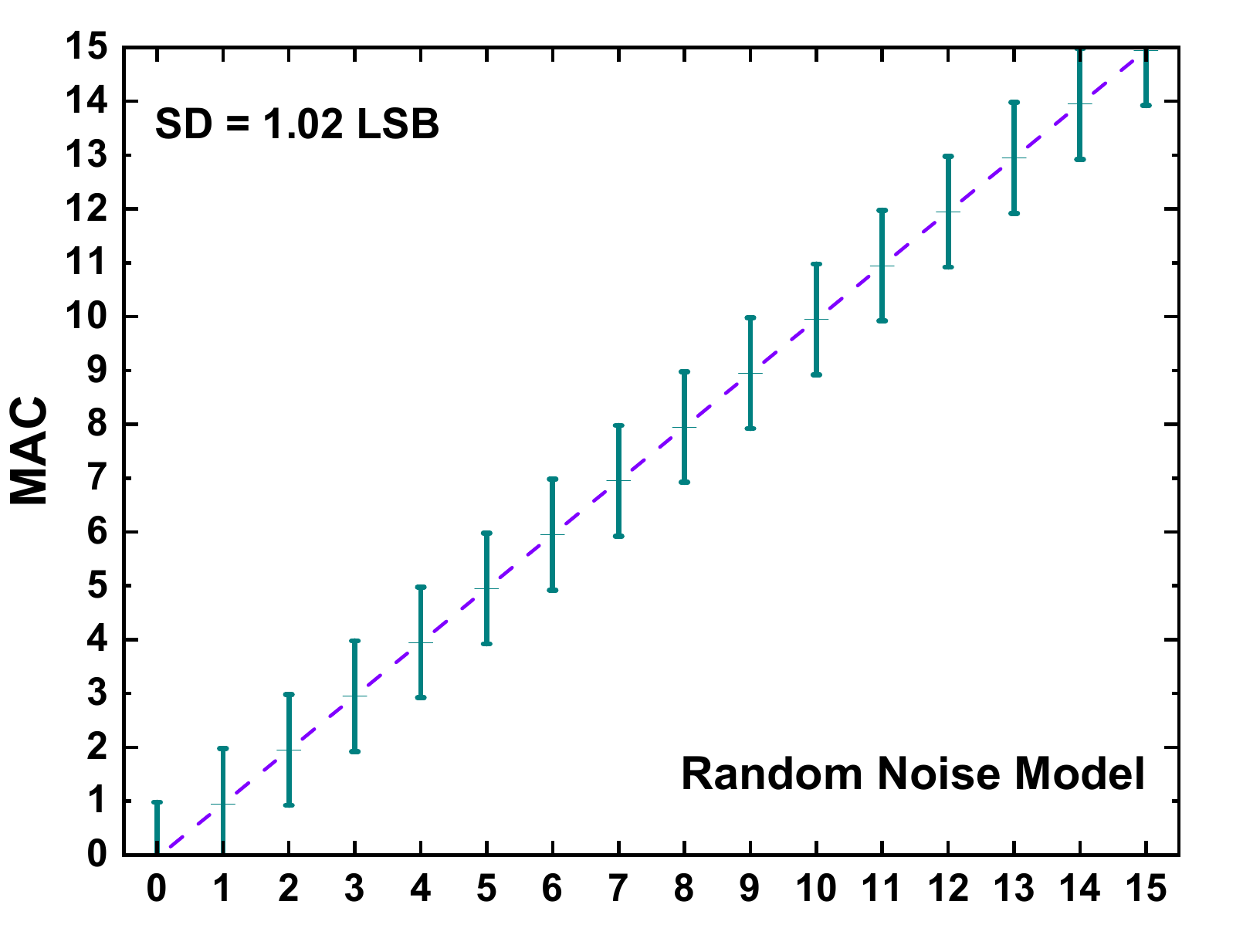}}
  \hfill
  \subfloat[]{\includegraphics[width=0.32\textwidth]{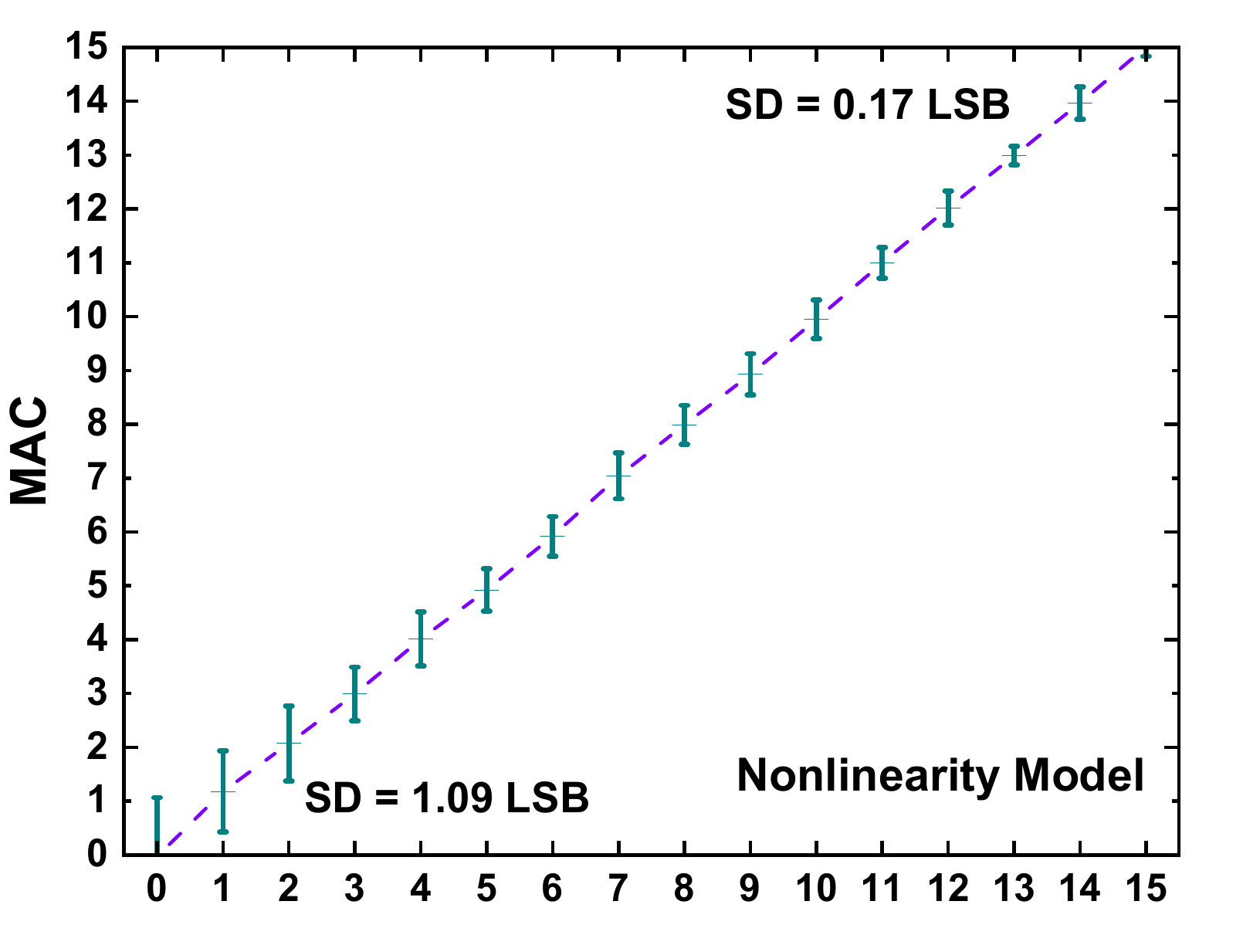}}
  \hfill
  \subfloat[]{\includegraphics[width=0.32\textwidth]{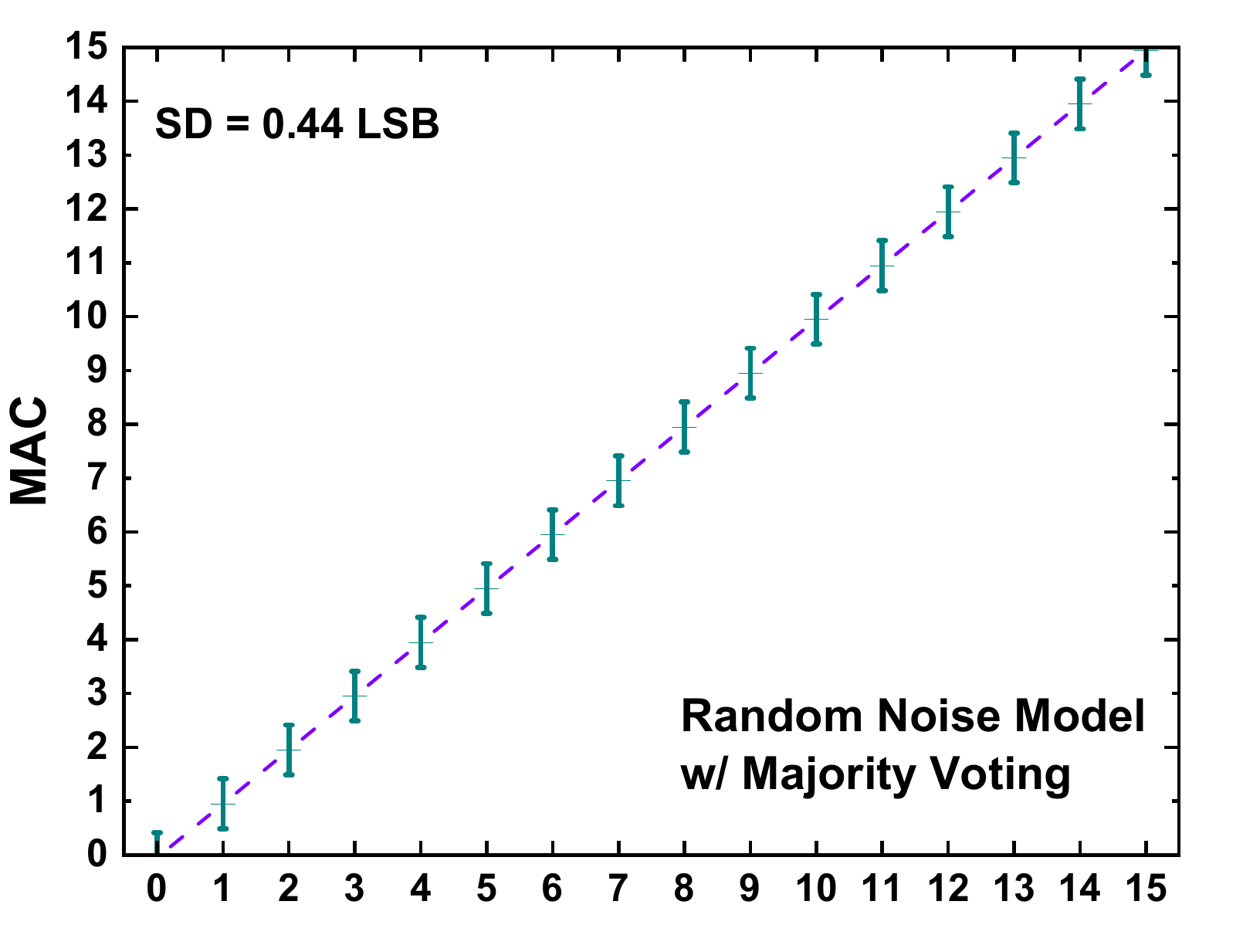}}
  \caption{Linearity plots of a 4-bit ADC under different noise models: (a) Random noise. (b) Nonlinearity-induced distortion. (c) Random noise with majority voting applied via 5 times oversampling.}
  \label{Fig_Linear_Plot}
\end{figure*}

To provide clearer insight into the noise models adopted in ASiM, we visualize the simulated noise behavior using linearity plots. Figs.~\ref{Fig_Linear_Plot}(a) and (b) show the MAC output characteristics of a 4-bit ADC under both the random noise model and the nonlinearity model. In both cases, the relative noise intensity is configured at $\text{LSB}_{\text{rms}} = 1.0$, and the error bars indicate the standard deviation ($\sigma$) of the MAC readout.

For the random noise model, each MAC output follows a Gaussian distribution with $\sigma \approx 1.02$ LSB. Under the nonlinearity model, smaller MAC values show larger variations ($\sigma \approx 1.09$ LSB), whereas the variation decreases gradually for larger MAC values ($\sigma \approx 0.17$ LSB). This trend is consistent with capacitor mismatch effects, where fewer active capacitors induce greater variation at lower MAC levels.

The majority voting approach mitigates analog noise by statistically reducing errors through oversampling. To illustrate this effect, we apply $5\times$ oversampling (with $\text{LSB}_{\text{rms}} = 1.0$) and show the corresponding linearity plot in Fig.~\ref{Fig_Linear_Plot}(c). As observed, majority voting reduces the standard deviation from approximately 1.02 LSB to 0.44 LSB, thereby improving MAC precision and enhancing inference accuracy.

\end{document}